\documentclass[traditabstract]{aa} % for the abstract without structuration 
\usepackage{graphicx}
\usepackage{txfonts}
\usepackage{subfigure}
\usepackage{color}
\usepackage{names}

\usepackage{natbib}
\bibpunct{(}{)}{;}{a}{}{,} % to follow the A&A style

\usepackage{multirow}
\usepackage{fullpage}

\def\eg{{\it e.g.} }

\def\ie{{\it i.e.} }
\def\Herschel{{\it Herschel } }

\begin{document}

\title{Scattering of small bodies by planets: a potential origin for exozodiacal dust ? }

    \author{Amy Bonsor
          \inst{1}
\and
           Jean-Charles Augereau \inst{1}
\and  
          Philippe Th\'{e}bault \inst{2}
          }
   \institute{UJF-Grenoble 1 / CNRS-INSU, Institut de Plan\'etologie et d'Astrophysique de Grenoble (IPAG) UMR 5274, Grenoble, F-38041, France \\     
                \email{amy.bonsor@gmail.com}
         \and
            LESIA-Observatoire de Paris, CNRS, UPMC Univ. Paris 06, Univ. Paris-Diderot, France\\}
           %  \email philippe.thebault@obspm.fr{}

   \date{Received ?, 20??; accepted ?, 20??}

%\abstract  {}{}{}{}{}
\abstract
{High levels of exozodiacal dust are observed around a growing number of main sequence stars. The origin of such dust is not clear, given that it has a short lifetime against both collisions and radiative forces. Even a collisional cascade with km-sized parent bodies, as suggested to explain outer debris discs, cannot survive sufficiently long. In this work we investigate whether the observed exozodiacal dust could originate from an outer planetesimal belt. We investigate the scattering processes in stable planetary systems in order to determine whether sufficient material could be scattered inwards in order to retain the exozodiacal dust at its currently observed levels. We use N-body simulations to investigate the efficiency of this scattering and its dependence on the architecture of the planetary system.  The results of these simulations can be used to assess the ability of hypothetical chains of planets to produce exozodi in observed systems. We find that for older ($>100$Myr) stars with exozodiacal dust, a massive, large radii ($> 20$AU) outer belt and a chain of tightly packed, low-mass planets would be required in order to retain the dust at its currently observed levels. This brings into question how many, if any, real systems possess such a contrived architecture and are therefore capable of scattering at sufficiently high rates to retain exozodi dust on long timescales.}% Our simulations suggest that the required high scattering rates are {\bf more likely to be }readily achievable if a planet is scattered into or migrates in to an outer belt, even for tens of Myr after the event. }

%This requires several planets, the outer planet of which sculpts the inner edge of the belt, scattering planetesimals, a small fraction of whom are scattered into the inner planetary system. We investigate the dependence on the architecture of the planetary system of the efficiency at which material is scattered inwards, in a manner that could be applied to many planetary systems. In terms of the older ($>100$Myr) observed exozodi, the dust could be retained at its current levels by material scattered inwards from a massive, large radii ($> 20$AU) and a chain of tightly packed, low-mass planets. We discuss the detectability of such outer belts. However, although our simulations show that there will be a low level flux of material scattered into the inner system in many planetary systems, our simulations suggest that the high levels of exozodiacal dust observed are more likely to result from material scattered inwards in the same manner, but in the aftermath (up to tens of Myr) of the scattering or migration of a planet into the outer belt. }

  \keywords{Planetary systems, Comets: general, Planets and satellites: dynamical evolution and stability, Planets and satellites: general, Zodiacal dust, Infrared: planetary systems, Methods: numerical
               }

   \maketitle

\section{Introduction}
\label{sec:intro}

Excess emission in the far-infrared or sub-mm, associated with cold, dusty debris discs is regularly observed around hundreds of main sequence stars (see \cite{wyattreview} for a review). There are now a growing number of stars with similar emission, but in the mid-infrared from material at higher temperatures. These observations have been associated with dust very close to the star, reminiscent of the solar system's exozodiacal cloud. There are now several dozen such sources discovered either using near or mid-infrared, interferometric observations with, for example, FLUOR/CHARA\citep[e.g.][]{Absil06, Absil08}, IONIC/IOTA\citep[e.g.][]{Defrere_Vega}, MIDI/VISIR \citep[e.g.][]{resolveHD69830}, or using their mid-infrared spectra with Spitzer \citep{Olofsson2012, Lisse12, Lisse2009}. Such sources have been compared to the solar system's exozodiacal cloud, but in general are several orders of magnitude brighter, for example $\eta$ Corvi is $1250\pm260$ times brighter than the solar system's exozodiacal dust cloud at $10\mu$m \citep{Millan-Gabet2011}.% others?. 
These observations can be modelled to determine estimates on the geometry, mass, position and grain properties. A well studied example is the 450Myr old A-star Vega \citep{Absil06,Absil08, Defrere_Vega}, where the best fit model found using Bayesian statistics to assess a grid of models, suggests that the emission is dominated by small grains ($<1\mu$m) in a $10^{-9} M_\oplus$ ring between $\sim 0.1-0.3 $AU \citep{Defrere_Vega}. More detailed compositional information can be determined for the handful of brighter sources detected by their Spitzer spectra in the mid-infrared, for example \cite{Olofsson2012, Lisse12}, amongst others. 

Once the potential alternative of a stellar companion is ruled out for the interferometric observations, the only explanation for this excess emission is emission from high levels of small dust grains. Such small grains, however, have a short lifetime, both against collisional processes (years) and radiative forces (years). If the observed discs are to survive for longer than year timescales, the small grains must be replenished. In outer debris discs, the observed small grains are thought to be continuously replenished in a steady-state collisional cascade from a population of unseen, large parent bodies. However, models for such collisional evolution (e.g. \cite{wyattdent02, wyatt99, wyatt07, colgrooming09, krivov08, lohne} amongst others), find that close to the star, the planetesimal belt required to supply the observed levels of small dust must be so dense that even kilometre sized parent bodies are rapidly destroyed by collisions \citep{wyatt07hot}. Thus, it is not possible for the observed small grains to be replenished by an insitu steady-state collisional cascade on long timescales (hundreds of years at 1AU). The suggestion has been made that the dust could be resupplied in a steady state collisional cascade if the population of bodies were on highly eccentric ($e>0.99$) orbits \citep{ecc_ring}. The origin of such a configuration is, however, difficult to conceive. Therefore, in general another explanation for the origin of the observed small dust is required, in particular for the older systems with exozodi.

Many suggestions have been made in the literature to explain these exozodi. The young systems ($\sim10-20$Myr) with hot dust are easier to explain and it seems probable that we are witnessing increased collision rates during terrestrial planet formation \citep{etatel, Smith2012,KB04}, whereas for the older systems, the observations are more puzzling. An obvious origin for the material is further out in the planetary system where it can survive on longer timescales \citep[e.g.][]{resolveHD69830, Lisse12, Absil06}. With the growing number of detections of planets and/or planetesimal belts, it seems that outer planetary systems are common. In fact some of the observed exozodis also have detections for outer planetesimal belts (e.g. Fomalhaut \cite{absil_fomb, Fom_herschel}, $\eta$ Corvi \cite{resolveHD69830, wyatt_etacorvi}, amongst many others). It remains, however, to be determined how this material is transported inwards and whether we are observing a transient or steady-state phenomena. 

Recent dynamical modelling of our solar system's exozodiacal dust cloud found good evidence that much of the observed dust originates from Jupiter Family Comets, scattered inwards from the Kuiper belt. Once inside Jupiter's orbit the dust's evolution is dominated by radiative forces, Poynting-Robertson-drag \citep{Nesvorny10}. Both scattering and PR-drag are transport processes that could occur in extra-solar planetary systems. Poynting-Robertson-drag is most influential in low luminosity systems, for example, it has been suggested to model observations of warm dust around Epsilon Eridani \citep{Eps_hotdust}. Scattering can occur both in a stable planetary system, such as the Jupiter Family Comets in our solar system, and at increased rates after a dynamical instability, such as the solar system's Late Heavy Bombardment. Evidence that we may be observing the aftermath of a collision between two large bodies exists from the presence of impact processed silica in the spectra of the hot dust \citep{Lisse12, Lisse2009}. Models of collisions between large bodies, such as the earth-moon forming collision, show that high levels of dust are produced \citep{Jackson2012}. Such collisions are by nature rare, in particular their probability decreases with the system age and thus, their probability must be compared to the growing numbers of systems with exozodiacal dust, in particular, the number of old (greater than Gyr) systems (\eg $\tau$ Ceti \citep{diFolco07}).

In this work we investigate whether sufficient material can be scattered inwards from an outer planetesimal belt, in a stable planetary system, to resupply an exozodi, or whether it is necessary to invoke further explanations. 
We consider the simple scenario where there are (several) planet(s) interior to an outer planetesimal belt and use N-body simulations to determine how efficiently they scatter material inwards from the outer belt. To assess the feasibility of this scenario, these efficiencies must be compared with observations of inner and outer belts. We start by discussing the details of our simulations in Sec.~\ref{sec:simulations}. Sec.\ref{sec:onepl} presents the illustrative example of single planet systems, which is expanded upon in Sec.\ref{sec:multi} with simulations to determine the scattering efficiencies in multi-planet systems. The implications of our results and their limitations are then discussed in Sec.~\ref{sec:discussion}, before conclusions are formed in Sec.~\ref{sec:conc}.

\section{Our simulations}
\label{sec:simulations}

%%%DEFINE STABLE PLANETARY SYSTEM!

The main purpose of this work is to investigate the rates at which material can be scattered from an outer planetesimal belt into the inner planetary system, with the view that it could contribute to the observed exozodiacal dust. This is necessarily dependant on the architecture of the planetary system; that is the orbits and masses of the planets, and the radial location and masses of the planetesimal belt and exozodi. We aim to constrain the planetary system architecture required in order to scatter material inwards at sufficient rates to produce the observed exozodiacal dust. We anticipate a high dependence of the scattering rates on the system architecture, thus, we hypothesis that the diversity of observed systems could potentially be explained by a diversity of planetary system architectures. For simplicity, we limit ourselves in this work to the scenario of an outer planetesimal belt. 

In order to assess the validity of this hypothesis, we use N-body simulations to determine the efficiency of scattering in planetary systems of different architectures. Given the huge diversity of planetary systems found by current observations\footnote{see \eg exoplanets.org}, it is necessary to limit our parameter exploration to a tiny region of the available parameter space, which is chosen such that the simulations can be used to make further deductions regarding the remaining parameter space. Our simulations determine the rate at which material is scattered inwards as a function of the total mass of material in the outer belt, that is the total mass in all bodies between small $\mu$m grains and km-sized bodies. The mass of the outer planetesimal belt can be determined, albeit in a model dependent manner, from observations of outer debris discs. For systems where an outer belt is not detected, this merely places an upper limit on the mass that could be hidden in an outer belt in this system. Using these masses it is possible to assess whether this hypothesis can be applied to real planetary systems with both outer planetesimal belts and exozodi. This requires a knowledge of the rate at which material needs to be supplied in order to sustain the observed levels of exozodiacal dust.

Observations of exozodiacal dust can be used, again in a model dependent manner, to determine a mass for the observed dust. It is clear that the observed dust is so small ($\mu$m) that it has a short lifetime, either against collisions or radiative forces, although the exact lifetime is difficult to calculate. Multiple epoch observations, separated by longer than the lifetime of the small dust, of a handful of systems \citep[e.g][]{Ciardi2001,Absil06,Absil08, Defrere_Vega},%, Defrere personal communication}, 
do not find significant variation in the observed fluxes. Therefore, it seems reasonable to make the assumption in this work that the small dust should be retained at its currently observed rates. It is, however, not clear whether this small dust is resupplied directly, or via collisions between somewhat larger (maybe only mm sized) grains also orbiting in the disc. In order to avoid a complicated discussion of the collisional evolution of material in the exozodi, we decide to make the base-line assumption that at least the currently observed mass of dust must be supplied to the exozodi, on the lifetime of this small dust. In general, this is likely to be (significantly) lower than the rate at which material with the size distribution of the outer belt (where the mass is dominated by large km sized bodies) needs to be scattered inwards, since it ignores the inefficient conversion of this material to small dust (discussed further in Sec.~\ref{sec:discussion}). However, if stable outer planetary systems cannot even scatter material inwards at this rate, then it is highly unlikely that they are capable of replenishing the exozodi. As our fiducial example we take the Vega system, where a best fit model finds $10^{-9}$M$_\oplus$ in between $0.2\mu$m and 1mm grains, with an estimated lifetime of 1yr \citep{Defrere_Vega}.  Thus, to retain Vega at its currently observed levels, the exozodi must be resupplied at a rate of at least $10^{-9}$M$_\oplus/$yr, although uncertainties in these values should be taken into account in the assessment of our simulations.

\subsection{The scattering process}
\label{sec:scattering}
%%%%%%%%%%%%%%%%%%%%%%%%%%%%%%%%%%

\begin{figure*}
\includegraphics[width=0.48\textwidth]{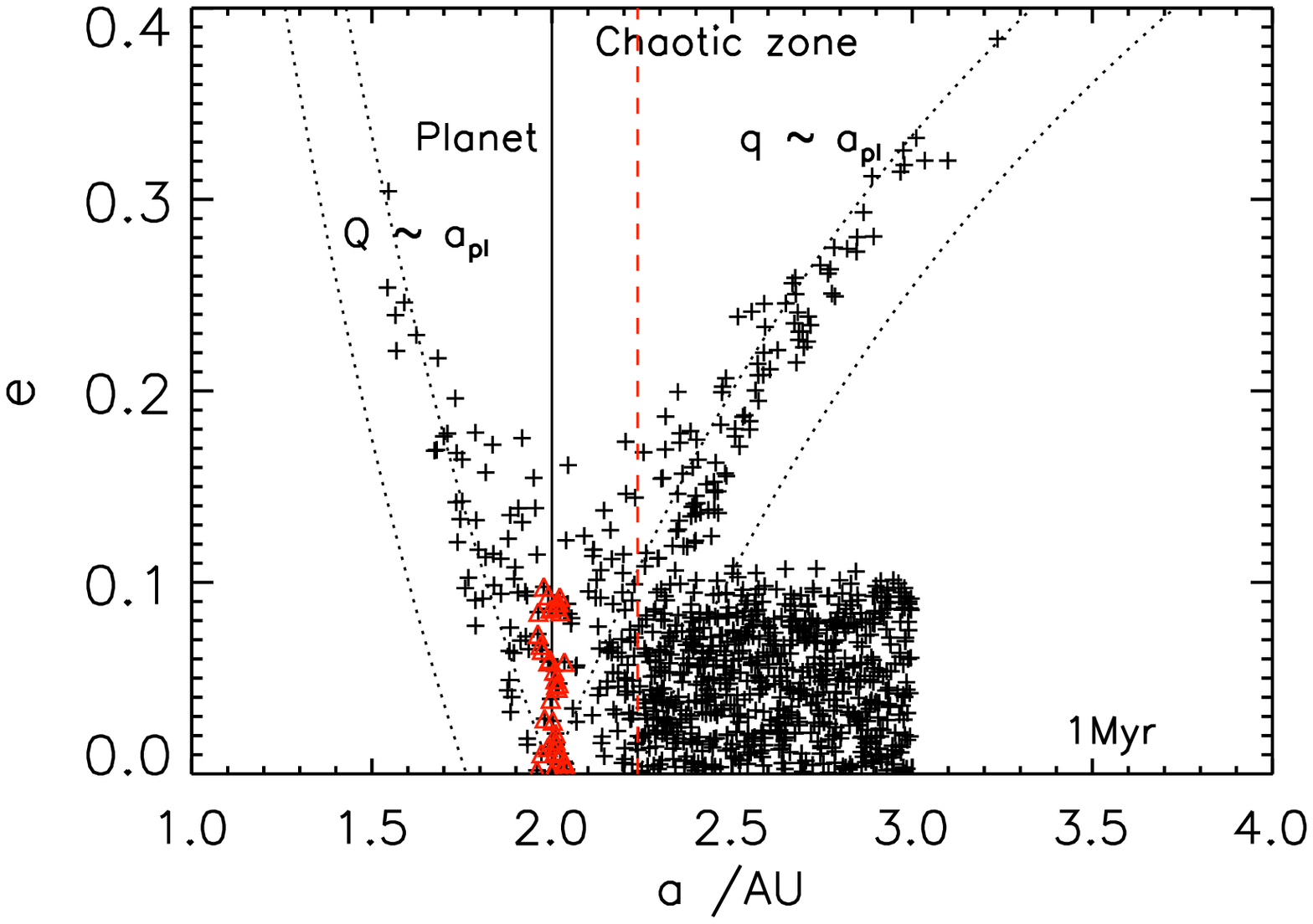}
\includegraphics[width=0.48\textwidth]{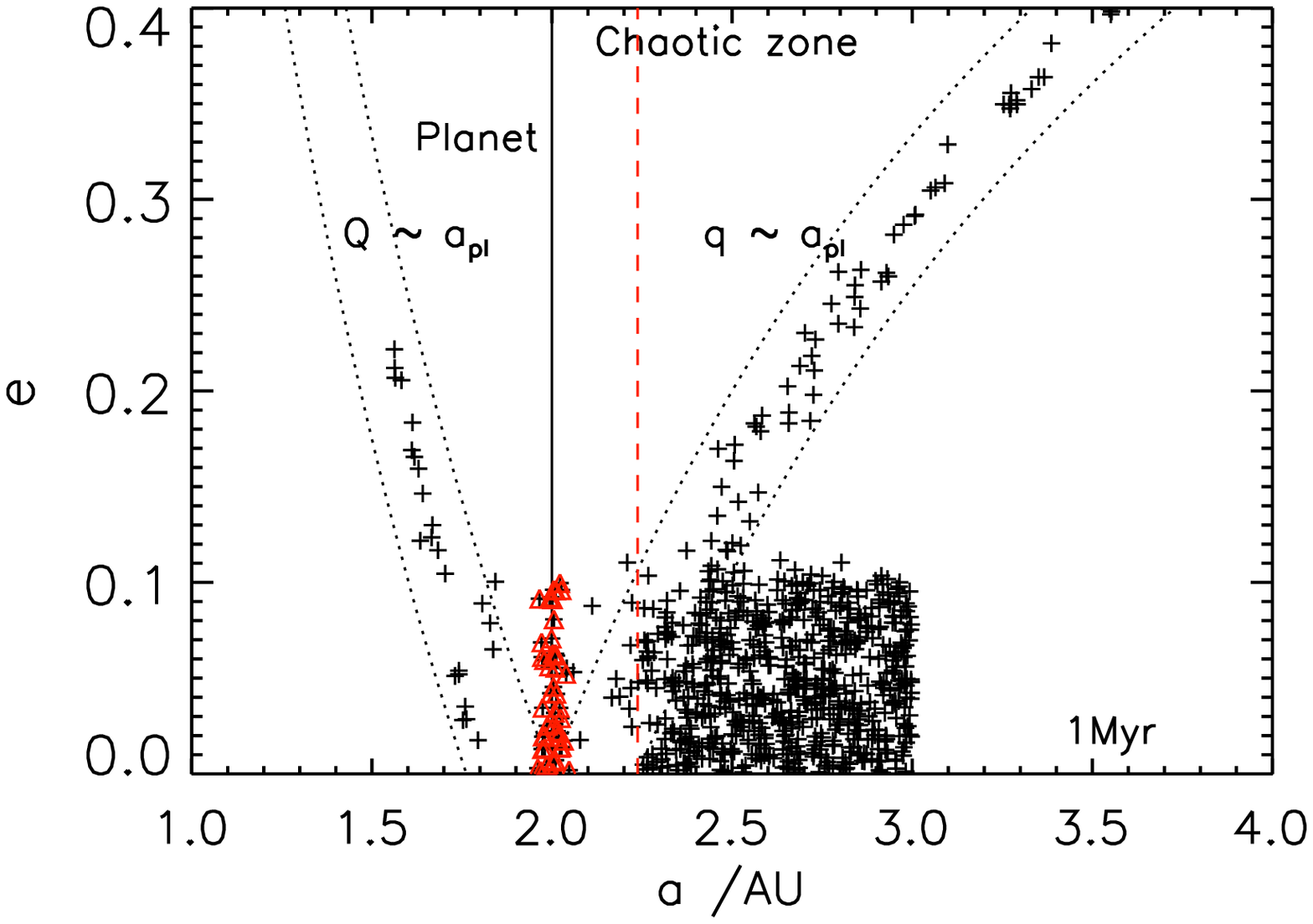}
\caption{The semi-major axes and eccentricities of particles scattered by a single Neptune mass planet on a circular orbit at 2AU around a $1M_\odot$ star, after 10,000yrs (left-hand panel) and 1Myr (right-hand panel). The set-up of the simulation is described in Sec.~\ref{sec:scattering}. Trojan's of the planet are labelled by red crosses. After 10,000yrs, only particles within the inner regions of the chaotic zone (Eq.~\ref{eq:chaos} with $C=2$, shown by the dashed lines) have been scattered, whilst by 1Myr, the full zone has been cleared. Particles are scattered onto orbits at constant pericentre ($q$) or constant apocentre ($Q$), thus lie between the dotted lines $q\sim a_{pl}$ and $q\sim a_{pl}- \delta a_{chaos}$  or $Q\sim a_{pl}$ and $Q\sim a_{pl} + \delta a_{chaos}$ ($Q\sim a_{pl}$). }
\label{fig:1mnep}
\end{figure*}

%%%%%%%%%%%%%%%%%%%%%%%%%%%%%%%%%%

%%%%%%%%%%%%%%%%%%%%%%%%%%%%%%%%%%

\begin{figure}
\includegraphics[width=0.48\textwidth]{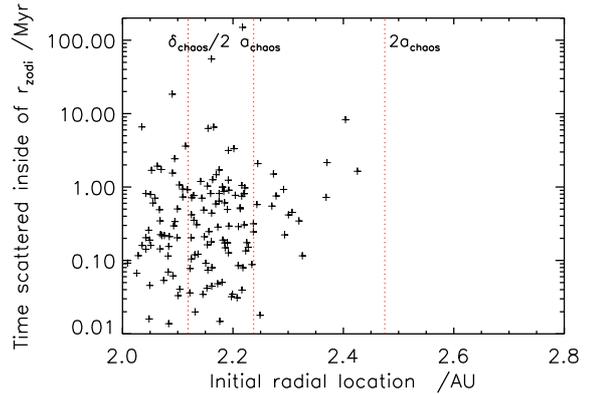}
\caption{The time at which particles are scattered inside of 1AU as a function of their initial semi-major axis, $a_i$, for the same simulation as shown in Fig.~\ref{fig:1mnep}. No particles that started outside of $a_i=2\delta a_{chaos}$ are scattered.  }
\label{fig:timecs}
\end{figure}

%%%%%%%%%%%%%%%%%%%%%%%%%%%%%%%%%%

In this work our focus is on the fate of planetesimals that are scattered during the planet's sculpting of the planetesimal belt. Of particular interest are those particles that are scattered into the inner planetary systems at late times, as these have the potential to replenish an exozodi.

The scattering of planetesimals by planets is a subject that has been investigated in much detail in the literature \citep[e.g.][]{Raymond09, Quillen2000, Beust96,Boley2012,bonsor_tiss, debesasteroidbelt}. %% Include solar system references too?!!
 Planets are often invoked to sculpt the inner edges of debris discs \citep[e.g.][]{Quillen06, chiang_fom, bonsor11}. A planet placed at the edge of a disc of planetesimals, as shown in Fig.~\ref{fig:1mnep}, will scatter particles that are close to it. These particles lie in a zone directly surrounding the planet, whose size can be estimated analytically by considering the overlap of mean motion resonances, given by \citep{Wisdom1980}:
\begin{equation}
\delta a_{chaos} = C \mu^{2/7} a_{pl},
\label{eq:chaos}
\end{equation}
where $\mu= \frac {M_{pl}}{M_*}$, the planet mass and $C$ is a constant, with various values between $C=1.3$ \citep{Wisdom1980}, $C=1.5$ \citep{DQT89, Quillen06} to $C=2$ \citep{chiang_fom} in the literature. In this work, we use the value $C=2$, as we find this is a good fit to our simulations (see Fig.~\ref{fig:1mnep}). Almost all particles in the planet's chaotic zone will be scattered, given sufficiently long timescales, however, some particles outside of this zone will also be scattered by the planet.

Although simulations such as \cite{Quillen06} show that planets generally clear their chaotic zone on less than 10,000 orbital periods, the effects of this clearing can continue on significantly longer timescales. Even in our own solar system, after Gyrs of evolution, there is evidence of Kuiper belt objects scattered by Neptune, for example, Jupiter Family Comets or Neptune's {\it scattered} disc. The best way to understand some of these processes is to consider a very simple example.

In Fig.~\ref{fig:1mnep}, we present the results of a simulation in which a $1M_{Nep}$ planet was placed on a circular orbit at 2AU, with an initially 'cold' ($e<0.1$ and $I<10^\circ$) planetesimal belt, extending from 2 to 3AU. Particles that start inside of the chaotic zone, and close to its edge, are scattered by the planet onto orbits of higher eccentricity, in general at either constant pericentre or apocentre. The scattering timescale is proportional to distance from the planet, thus the chaotic zone is cleared from inside outwards. In the left-hand panel, it is clear that after 10,000yrs of simulation, the chaotic zone is still being cleared; all particles initially in the inner regions have been scattered, whilst there are still some particles on unperturbed orbits in the outer regions of the chaotic zone. On the other hand, as shown in the right-hand panel, after 1Myr of simulation, the full chaotic zone has been cleared. Particles are scattered onto orbits at constant pericentre or apocentre, such that the scattered particles lie between the lines $q\sim a_{pl}$ and $q\sim a_{pl}- \delta a_{chaos}$  or $Q\sim a_{pl}$ and $Q\sim a_{pl} + \delta a_{chaos}$, where they are scattered multiple times, until eventually they are either ejected, or hit the planet or star. This can be on significantly longer timescales than the clearing of the chaotic zone. Such particles are equivalent to Neptune's {\it scattered} disc in our solar system.

In terms of the formation or replenishment of exozodi, we are interested in particles that are scattered close to the star, inside of a limiting radius that we call $r_{zodi}$. Whether this refers to the location of the exozodi or the radius at which Poynting-Robertson or other drag forces dominate is irrelevant to the current simulations. Particles that are scattered inside of this radius originated in (or near to) the planet's chaotic zone, but have been scattered multiple times by the planet, generally evolving at constant apocentre, until their eccentricity is sufficiently high that they are scattered inside of $r_{zodi}$. Thus, the total amount of material scattered generally depends on the size of the chaotic zone and critically the planet's mass. 

Fig.~\ref{fig:timecs} shows the time at which particles are scattered inside of $r_{zodi}=1$AU, by the same $1M_{Nep}$ planet on a circular orbit at 2AU, as a function of their initial position. The broad trend is for an increase in the time at which particles reach $r_{zodi}=1$AU with increasing initial semi-major axis. This was anticipated, as the planet scatters particles close to it more quickly. This trend is, however, very smeared out, in a way that would not be the case if we merely plotted the time at which the particles were initially scattered. This is because once particles have been scattered a few times by the planet, their further scattering looses its dependence on the particle's initial orbit. Given that most particles are scattered many times before they end up inside of $r_{zodi}$, the trend in Fig.~\ref{fig:timecs} is somewhat blurred.

Critically, Fig.~\ref{fig:timecs} shows that the scattering rates, even at late times, can be influenced by particles that started on orbits very close to the star. This is unfortunate since it is not clear whether or not particles are expected on such orbits after planet formation. As discussed in further detail in Sec.~\ref{sec:initialconditions}, the end conditions of planet formation are not well constrained. It is, however, clear that planets clear a zone directly surrounding them, although there are no precise constraints on the size of this zone, which may well differ between terrestrial planet and gas giant formation. It is not the purpose of this work to investigate the end conditions of planet formation, nor to rely greatly upon their outcome, therefore, we restrict the initial semi-major axes of our test particles to $a_{min}$, where $a_{min}>a_{pl}$. This removes particles from very close to the planet, that would have almost definitely been accreted onto the planet, but ensures that we include particles close the edge of the planet's chaotic zone. Given that our initial test simulations show that such particles are only scattered on timescales greater than $10,000 T_{pl}$, it seems unlikely that all of these particles would have been significantly influenced, by the smaller planetary core, to have been accreted during planet formation. Thus, although the value of $a_{min}$ may be somewhat arbitrarily chosen, it should nonetheless provide a good representation of the behaviour anticipated in a real planetary system.  This, however, ignores the effect of migration during planet formation, that could potentially leave gaps in the planetary system. These affects and the influence of our choice of $a_{min}$ are discussed further in Sec.~\ref{sec:initialconditions}.

%%%%%%%%%%%%%%%%%%%%%%%%%%%%%%%%%%

\begin{figure}

\includegraphics[width=0.48\textwidth]{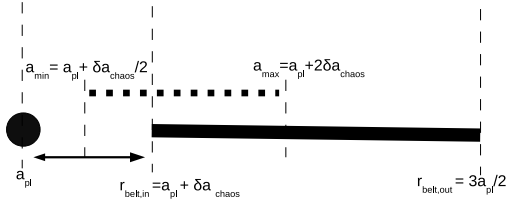}
\caption{A diagram to illustrate the set-up of our simulations. The large circle represents the planet, at $a_{pl}$. The real belt, at late times, is shown by the thick line between $r_{belt,in} =  a_{pl,out}+\delta a_{chaos}$ to $r_{belt,out}=\frac{3 r_{belt,in}}{2}$. To simulate this belt we need only consider test particles that start between $a_{min} = a_{pl,out}+ \frac{\delta a_{chaos}}{2}$ and  $a_{max} = a_{pl,out}+ 2\delta a_{chaos}$, represented by the dotted, thick line. }
\label{fig:diag}
\end{figure}

%%%%%%%%%%%%%%%%%%%%%%%%%%%%

\subsection{Details of the simulations}
\label{sec:setup}
Our simulations are run using the RMVS3 integrator as part of the SWIFT package \citep{swift_rmvs}. This has an adaptive timestep. We consider planetary systems that have an outer planetesimal belt, running from $r_{belt,in}$ to $r_{belt,out}$, and a total mass of $M_{belt}$ in all bodies (between the blow-out size,  $D_{bl}$ and a maximum size, $D_{max}$, as in Eq.~\ref{eq:mmax}). The planets are on circular, coplanar orbits, interior to this outer belt. We consider that the outer planet, orbiting with a semi-major axis of $a_{pl,out}$ has sculpted the inner edge of the outer belt (see for example Fig.~\ref{fig:1mnep}). The outer planet is therefore placed interior to the outer belt's inner edge by its chaotic zone (Eq.~\ref{eq:chaos}), \ie $a_{pl,out}=r_{belt,in}- \delta a_{chaos}$. For our initial simulations we only consider a single planet, whilst later in this work (Sec.~\ref{sec:multi}) we consider chains of several equal mass planets, of mass $M_{pl}$, separated equally in units of their mutual Hill's radii, given by \citep[e.g.][]{Raymond09, Chambers96}
\begin{equation}
R_H = \frac{(a_1 + a_2)}{2}\left(\frac{M_1 + M_2}{3M_*}\right)^{1/3},
\end{equation}
where $a_1$ and $a_2$, $M_1$ and $M_2$ are the semi-major axes and masses of the two planets. 

Our outer planetesimal belt is composed of $N$ test particles, whose initial semi-major axes are randomly selected between $a_{min}$ and $a_{max}$, initial eccentricities between 0 and $e_{max}=0.1$, initial inclinations between 0 and $I_{max}=10^\circ$, and whose initial longitude of ascending node, argument of pericentre and mean anomalies are between 0 and $2\pi$. Although we consider our simulations representative of a belt between $r_{belt,in}$ and $r_{belt,out}$, it is not necessary to simulate particles covering the full belt (\ie $a_{min}= r_{belt,in}$ and $a_{max}=r_{belt,out}$). Initial test simulations (see Fig.~\ref{fig:timecs}) found that a single planet has little to no influence on particles outside of twice the width of its chaotic zone. Thus, to reduce computing time, we fix $a_{max}$ at $a_{pl,out}+ 2\delta a_{chaos}$. As will be discussed in further detail in Sec.~\ref{sec:initialconditions}, the planet will already have cleared a zone surrounding it during its formation. The exact size of this zone is not well constrained and particles this close to the planet are scattered on average on shorter timescales, therefore we arbitrarily fix $a_{min}$ at $a_{pl,out}+ \frac{\delta a_{chaos}}{2}$, although the dependence of our simulations on this value should be noted. 

The results of our simulations are scaled to the full belt of width $r_{belt,in}$ to $r_{belt,out}$. Care must be taken when referring to the mass of the outer belt. In this work we refer to the total mass of the outer belt as the total mass in a belt between $r_{belt,in} =  a_{pl,out}+\delta a_{chaos}$ and $r_{belt,out}=\frac{3 r_{belt,in}}{2}$, except for in the examples where the outer belt radial extent is known from observations. This outer radius for the belt is an entirely arbitrary choice, and our results can be readily scaled to any value. It, however, enables a comparison between examples with different outer planet masses to be readily made. 

The details of the set-up used in our simulations are shown in Fig.~\ref{fig:diag} and all the important parameters are listed in Table~\ref{tab:params}. In order to avoid unnecessary computing, we consider that particles have left the planetary system if their eccentricity increases above 1 or they are scattered outside of $r_{max}$, which we take as 10AU in the one planet simulations and 1000AU in the multi-planet simulations. %% CHECK!  

To determine the rate at which material is scattered inwards, we track the rate at which particles are scattered inside of $r_{zodi}$, the nominal location of the exozodi, which we generally take to be 1AU. Given the necessity to limit the number of test particles in our simulations, we consider that each test particle represents a fraction of the mass in the outer belt, contained in a suitable number of particles that reflect the size distribution of the outer belt. The fraction of the outer belt mass represented by each test particle is assigned based on the particles initial semi-major axis, assuming a mass surface density of the belt given by $\Sigma (r) dr \propto r^{-\alpha} dr$, where $\alpha=1$, and that $a\sim r$ (valid for the low initial particle eccentricities used here). Given the potential difference between the simulated belt and the actual belt (see Fig.~\ref{fig:diag}), each particle is assigned a fraction of the total mass in the belt, extending between $r_{belt,in}$ and $r_{belt, out}$:  
\begin{equation}
m_{part}=\left(\frac{1}{N_{pp}}\right) \left( \frac{ r_{belt,out}-r_{belt, in}}{a_{max}-a_{min}}\right), 
\label{eq:mpart}
\end{equation}
where $N_{pp}$ is the number of test particles placed between $a_{min}$ and $a_{max}$ at $t=0$. Using this formulation, this rate can readily be scaled to an outer belt of a different width (\ie with a different $r_{belt,out}$. The rate at which material is scattered inwards is then calculated from the fraction of the belt mass assigned to each particle divided by the time between adjacent collisions.

%%%%%%%%%%%%%%%%%%%%%%%%%%%%%%%%%%
%% A TABLE WITH ALL OF THE PARAMETERS USED IN THIS WORJK!
%\begin{minipage}[c]{2\textwidth}
\begin{table*}
\begin{center}

\caption{A list of the parameters used in this work}

\begin{tabular}{l l l}
        
       \hline
Parameter & Nominal value in this work &Description \\ \hline

$\Delta$ & & The separation of the planets, in units of their mutual Hill's radii\\
$\delta a_{chaos}$ & & The radial extent of the chaotic zone in units of the planet's semi-major axis \\
$\rho$ &$2700kg m^{-3}$ & The density of the individual particles in the outer belt \\
$a_{i}$ & & The initial semi-major axis of a particle, before it is scattered by a planet\\

$a_{pl}$ & & The planet's semi-major axis \\
$a_{pl,out}$&$r_{belt,in}- \delta a_{chaos}$ & The outer planet's semi-major axis \\
$a_{min}$ &$a_{pl,out}+ \frac{\delta a_{chaos}}{2}$ &The minimum initial semi-major axis of test particles in this simulations \\
$a_{max}$& $a_{pl,out}+ 2\delta a_{chaos}$ &The maximum initial semi-major axis of test particles in this simulations \\
$a_{fill}$ & 5AU & Planets are placed between this radius and the outer belt. \\

$D_{max}$& 60km&  The size of the largest body in the outer disc \\
$dr$ & $\frac{dr}{r}=2$ & The width of the outer belt \\ 
$e_{max}$& 0.1&The maximum initial eccentricity of test particles in this simulations \\
$e_i$ & & The eccentricity of a particle before it is scattered by the planet\\

$F_{disc}$ &  & The disc flux \\
$F_{disc}$ &  & The stellar flux \\

 $\langle e \rangle$ &0.05& The average eccentricity of particles in the outer belt \\

$I_{max}$ & $10^\circ$&The maximum initial inclination of test particles in this simulations \\

$M_*$ & $1M_\odot$ unless otherwise specified & the stellar mass\\
 $M_{det}$& & The minimum mass of a planetesimal belt such that it can be detected \\
$M_{max}$& & The maximum mass of a collisionally evolving planetesimal belt \\

$M_{part}$&  & The fraction of the total outer belt mass assigned to each test particle\\
$M_{pl}$& & The planet mass \\
$N_{pp}$ & & The number of particles used in the simulation\\
$Q_D^*$ & $150Jkg^{-1}$& The dispersal threshold\\

$q_{min}$ & & The minimum pericentre of a particle, with a Tisserand parameter of $T_i$, \\
& & scattered by a single planet on a circular orbit \\

$R_H$ & & Hill's radii \\
$r$ & & The central radius of the outer belt \\
$r_{belt,in}$ &$a_{pl,out}+ \delta a_{chaos}$  &The inner radius of the outer belt. \\
$r_{belt,out}$ &$\frac{3 r_{belt,in}}{2}$ &The outer radius of the outer belt.  \\
$r_{zodi}$ & 1AU & The position of the exozodi, or the inner radial limit within which most \\
& & scattered material will go on to produce an exozodi\\
$r_{max} $ & 10 or 100 AU& The distance outside of which particles are considered removed from the simulations \\

$r_{min} $ & & The minimum radius of a planetesimal belt in order that a given mass of material can survive \\
& & against collisions for the age of the system\\
$r_{belt,min} $ & & The minimum radius required by a planetesimal belt such that it can scatter material inwards at \\
 & & a sufficiently high rate to retain the observed levels of exozodiacal dust \\

$T_i$ &  & The initial value of the Tisserand parameter \\

%$A$ &  & Parameters fitted to the rate at which particles are scattered inwards\\
%$\gamma$ & &Parameters fitted to the rate at which particles are scattered inwards \\
    \hline

\end{tabular}
\end{center}
\label{tab:params}
\end{table*}          

%%%%%%%%%%%%%%%%%%%%%%%%%%%%%%%%%%%%%%%%%%%%%%%%%%%%%%%%%%%%%%%%%%%%%%%%%%%%%%%%%%%%%%%%%%%%%%%%%%%%%%%%%%%%%%%%%%%%%%%%%%%%%%%%%%%%%%%%%%%%%%%%%%%%%%%
%%%%%%%%%%%%%%%%%%%%%%%%%%%%%%%%%%%%%%%%%%%%%%%%%%%%%%%%%%%%%%%%%%%%%%%%%%%%%%%%%%%%%%%%%%%%%%%%%%%%%%%%%%%%%%%%%%%%%%%%%%%%%%%%%%%%%%%%%%%%%%%%%%%%%%%

%%%%%%%%%%%%%%%%%%%%%%%%%%%%%%%%%%

\begin{figure}

\includegraphics[width=0.48\textwidth]{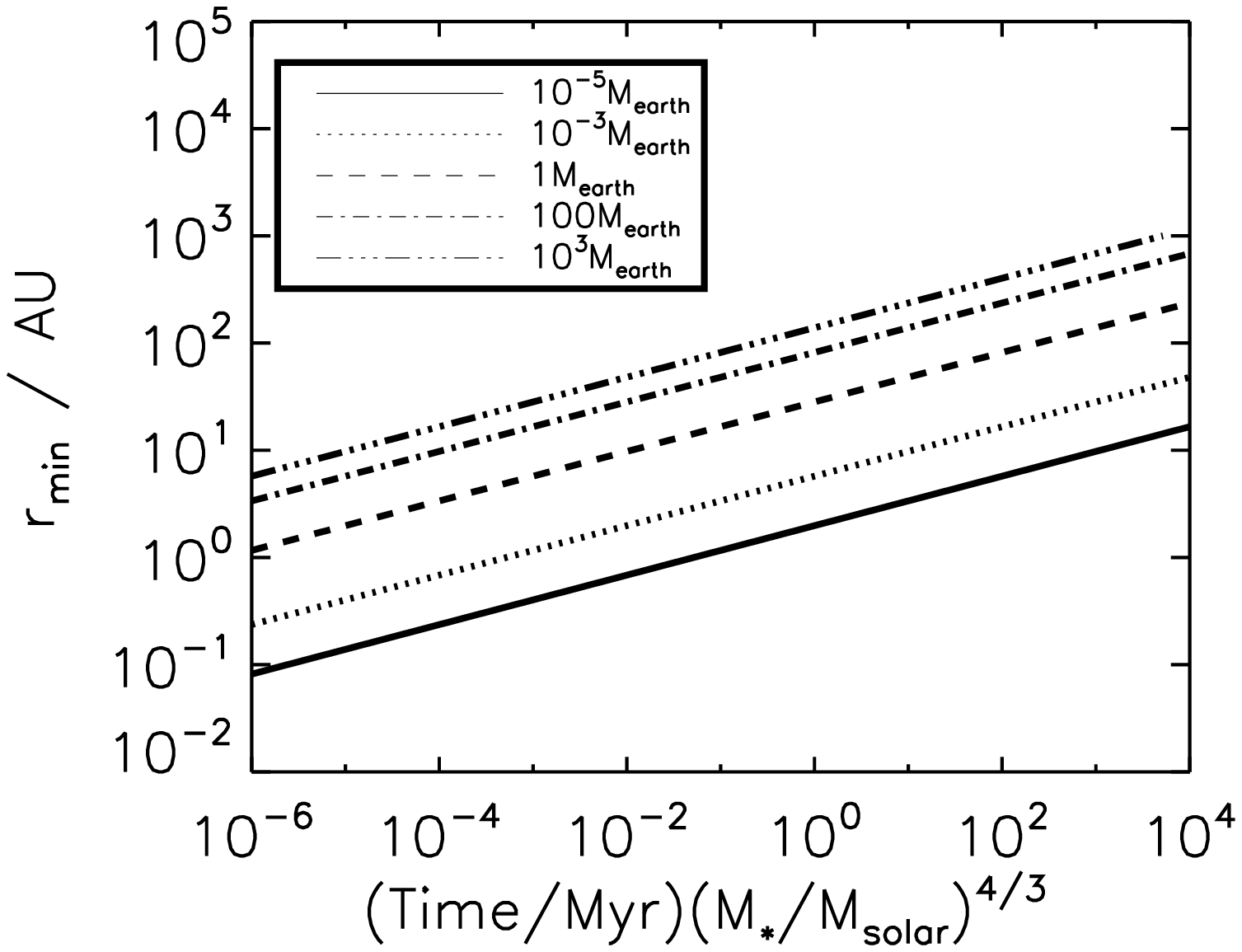}
\caption{The minimum central radius of a planetesimal belt, such that a total mass of either $M_{disc}=10^{-5}$, $10^{-3}$, $1$, or $10M_\oplus$, can survive, as a function of age. This is calculated using Eq.~\ref{eq:rmin}, assuming a steady-state collisional equilibrium, with bodies up to a maximum size $D_{max}=60$km and fiducial values for the other parameters taken from fits to observations of main sequence debris discs, quoted in Sec.~\ref{sec:mass}. The results scale directly with $\frac{D_{max}}{M_{disc}}$ and the time axis can be scaled by a factor involving the stellar mass, $\frac{M_*}{M_\odot}^{4/3}$.}% This plot was made using Eq.~\ref{eq:mmax}, taking the fiducial values for the other parameters stated in the text.  }%[dr/r=0.5- change?]  }
\label{fig:rmin}
\end{figure}

%%%%%%%%%%%%%%%%%%%%%%%%%%%%%%%%%%

\subsection{The survival timescale of massive outer belts}
\label{sec:mass}
As can be logically expected, and as will be quantitatively investigated in our study, the scattering of bodies from an outer belt to an exozodi is facilitated for massive outer belts. There is, however, a theoretical limit as to how massive such a belt can be, imposed by its collisional evolution. This concept of a maximum possible belt mass for a given system's age is a very important constraint when discussing our results.  It is therefore important to introduce it before presenting and discussing our simulations.

Debris discs are collisional system in which larger bodies are continuously collisionally ground down to smaller bodies and the smallest bodies are removed by radiation pressure. Thus, over time the total mass in the disc is depleted. Care must be taken in referring to the total disc mass, which here refers to the total mass of all bodies in the collisional cascade,  extending from the observed small dust grains ($\mu$m) to km sized bodies. There may be larger bodies that are not yet in collisional equilibrium, it is however, impossible to obtain any information about such bodies from the observations and we therefore choose to disregard them in this work. Models for the collisional evolution of such discs from \cite{wyatt99, wyattres,wyatt07hot,wyatt07} find that there exists a maximum total disc mass that can survive against collisions on long timescales. This maximum mass as a function of the disc radius, $r$, is given by \citep{wyatt07hot}:
\begin{eqnarray}
&M_{max} (t)= 5.2\times 10^{-13}\; {\mathrm M}_{\oplus} \times \\\nonumber
& (\rho/kgm^{-3}) \; (r/\mathrm{AU})^{13/3}\; (\frac{dr}{r}) (Q_D^*/\mathrm{Jkg}^{-1})^{ 5/6}\;(D_{max}/km)\\\nonumber 
&(M_*/M_{\odot})^{-4/3})\;\langle e \rangle^{-5/3}\; (t/\mathrm{Myr})^{-1} ,
\label{eq:mmax}
\end{eqnarray}

where $\rho$ is the bulk density of the bodies, $r$ the central radius of the belt of width $dr$, $Q_D^*$ the dispersal threshold, $D_{max}$ the size of the largest body in the disc, $M_*$ the stellar mass, $\langle e \rangle$ is the average eccentricity of particles in the disc, assumed to be constant and $t$ is the age of the system, or strictly the time after the disc was stirred. Written in this form, it was assumed that the disc contains a Dohnanyi size distribution of bodies \citep{Dohnanyi} with a power index of $\frac{7}{2}$ \citep{Tanaka96} between the smallest size body that is not blown out of the system by radiation pressure and the significantly larger maximum size for bodies in the disc, $D_{max}$. The mass is dominated by the largest bodies. It also assumes that particles can be catastrophically destroyed by particles significantly smaller than themselves, that the disc is thin (of width $dr$) and does not spread radially.

This maximum mass is a strong function of the disc radius. We can use it to determine the minimum radius of a planetesimal belt in which a particular mass can survive against collisions, up to a given age. Thus, the minimum radius at which a disc of mass, $M_{disc}$, can survive is
\begin{eqnarray}
&r_{min} = 6.8 \times 10^2 \; AU \;\times \\\nonumber
& (M_{disc} /M_{\oplus})^{3/13} (M_*/M_{\odot})^{4/13}\langle e \rangle^{5/13} (t/Myr)^{3/13}\\\nonumber
&(D_{max}/km)^{-3/13} (\rho/kgm^{-3})^{-3/13} (\frac{dr}{r})^{-3/13} (Q_D^*/Jkg^{-1})^{-5/26}.
\label{eq:rmin}
\end{eqnarray}

This minimum radius is particularly interesting because it enables us to place a radial limit on the position of an outer belt if sufficient mass is to survive for a given age. This minimum radius is plotted in Fig.~\ref{fig:rmin} as a function of time. Formally this is the time since the collisional velocities in the disc reached high enough values to be catastrophic and the system was stirred, which we assume to be approximately equal to the age of the system. We take fiducial values for the model parameters that come from fitting observations of debris discs around main sequence A stars \citep{wyatt07}. Thus, the maximum body is $D_{max}=60$km, eccentricity $\langle e \rangle=0.05$, the density $\rho=2700$kgm$^{⁻3}$ and the width $\frac{dr}{r}=0.5$. In order to allow some of these parameters to be easily varied, the time is plotted multiplied by a factor involving the stellar mass and the ratio of the disc mass to the maximum diameter particle is fixed, rather than the exact disc mass. Planetesimal belts in steady state can exist above and to the left of the minimum lines plotted. 

Fig.~\ref{fig:rmin} confirms our earlier assertion that exozodi cannot be supplied by a steady-state collisional cascade of massive, unseen, parent bodies orbiting at their radial location. Consider the example of Vega, where the best fit to the observations of the exozodi find $1.9 \times 10^{-9} M_{\oplus}$ of material, in 0.2$\mu$m to 1mm sized bodies \citep{Defrere_Vega}. If this material were to be supplied in a steady-state size collisional cascade from unseen bodies extending up to $D_{max}=60$km in size, at the same radial location (\ie inside of 1AU), an absolute minimum on the total disc mass required would be $10^{-5}$M$_{\oplus}$. Fig.~\ref{fig:rmin} clearly shows that such a mass of material cannot survive at 1AU for longer than $\sim30$Myr, less than 10\% the age of Vega. However, such a mass (and significantly more) can survive for $455$Myr age of Vega \citep{Yoon2010} \footnote{although note the factor of $(2.1M_\oplus)^{4/3}$ necessary in Fig.~\ref{fig:rmin}  to account for the mass of Vega \citep{Yoon2010}}, in a belt outside of 60AU  (the observed debris disc radius \citep{muller}). 

%%%%%%%%%%%%%%%%%%%%%%%%%%%%%%%%%%%%%%%%%%%%%%%%%%%%%%%%%%%%%%%%%%%%%%%%%%%%%%%%%%%%%%%%%%%%%%%%%%%%%%%%%%%%%%%%%%%%%%%%%%%%%%%%%%%%%%%%%%%%%%%%%%%%%%%
%%%%%%%%%%%%%%%%%%%%%%%%%%%%%%%%%%%%%%%%%%%%%%%%%%%%%%%%%%%%%%%%%%%%%%%%%%%%%%%%%%%%%%%%%%%%%%%%%%%%%%%%%%%%%%%%%%%%%%%%%%%%%%%%%%%%%%%%%%%%%%%%%%%%%%%
%%%%%%%%%%%%%%%%%%%%%%%%%%%%%%%%%%%%%%%%%%%%%%%%%%%%%%%%%%%%%%%%%%%%%%%%%%%%%%%%%%%%%%%%%%%%%%%%%%%%%%%%%%%%%%%%%%%%%%%%%%%%%%%%%%%%%%%%%%%%%%%%%%%%%%%

\section{One planet}
\label{sec:onepl}
The simplest possible planetary system is one containing an outer planetesimal belt and a single interior planet on a circular orbit.  Although possibly not a likely structure for many real planetary systems, this provides a simple to understand, illustrative example of the scattering processes that occur in more complicated multi-planet systems, which we defer to later in this work. We investigate the rate at which particles are scattered to the position of the exozodi. However, a single planet is not capable of scattering particles infinitely far and therefore it is useful to firstly consider analytically the position of the planet relative to the outer belt and the exozodi.

%%%%%%%%%%%%%%%%%%%%%%%%%%%%%%%%%%%%%%%%%%%%%%%%%%%%%%%%%%%%%%%%%%%%%%%%%%%%%%%%%%%%%%%%%%%%%%%%%%%%%%%%%%%%%%%%%%%%%%%%%%%%%%%%%%%%%%%%%%%%%%%%%%%%%%%

\subsection{An analytical limit on the maximum semi-major axis of the planet }
\label{sec:tiss}
If particles scattered by a planet are to produce an exozodi, the planet must be capable of scattering them sufficiently far in, to $r_{zodi}$. If we consider planets on circular orbits, the conservation of the Tisserand parameter places an analytical limit on how far a planet can scatter a particle \citep{bonsor_tiss}. This can be expressed as a minimum allowed pericentre for the particle's orbit, given by (Eq.3 of \cite{bonsor_tiss}) \begin{equation}
q_{min} (T_i) = a_{pl} \frac{-T_i^2 +2T_i +4 - 4 \sqrt {3-T_i}}{T_i^2-8},
\label{eq:qmin}
\end{equation} where $T_i$ is the initial value of the Tisserand parameter, for a particle with semi-major axis, $a_i$, eccentricity, $e_i$ and inclination, $I_i$ (Eq.3 of  \citep{bonsor_tiss}). This limit is only strictly valid for low mass planets on circular orbits, where long range interactions are rare. It, however, provides a good approximation to the behaviour even of Jupiter mass planets on mildly eccentric orbits, although there may be a small flux of particles scattered inside of this minimum pericentre. 

\begin{figure} 
\includegraphics[width=0.48\textwidth]{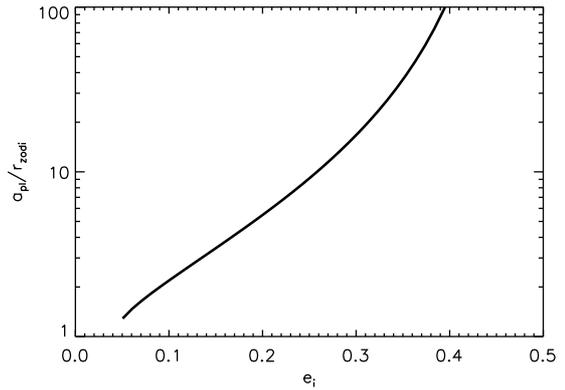}
\caption{An analytical limit on the maximum semi-major axis of a planet, if it is to be able to scatter particles as far inwards as the position of the exozodi at $r_{zodi}$, as a function of the particle's initial eccentricity. This is calculated using Eq.~\ref{eq:qmin}, and the particle's initial eccentricity is a proxy for the Tisserand parameter value, as described in Sec.~\ref{sec:tiss}. This limit only strictly exists for planets on circular orbits and vanishes for $e_i>0.46$ (or $T_i<2$), or for very low eccentricity particles that do not cross the planet's orbit ($e_i>0.04$ or $T_i<3$).  }
   \label{fig:tiss_limit}
\end{figure}

This minimum pericentre can be used to determine whether or not a particular configuration of planet and planetesimal belt can scatter any material to the position of the exozodi, $r_{zodi}$. For a single planet it provides a limit on the maximum semi-major axis of the planet. This depends on the initial orbital parameters (Tisserand parameter value) of the particles before they are scattered by the planet and is determined by equating $q_{min}(T_i)= r_{zodi}$. This maximum semi-major axis is plotted in Fig.~\ref{fig:tiss_limit} as a function of the particle's Tisserand parameter value, except that the particle's initial eccentricity is used as a proxy for its value of the Tisserand parameter. This means assuming values for the particle's initial inclination of $I_i\sim 2e_i$, and initial semi-major axis of $a_i= a_{pl}+\delta a_{chaos}(1M_{Nep})$, which places the particle initially at the edge of the planet's chaotic zone. In this plot, we consider a Neptune mass planet, although the dependence on the planet mass of the results is negligible. In terms of our single planet simulations, given that we start particles in a fairly cold belt, with $e_{max}=0.1$, Fig.~\ref{fig:tiss_limit} shows that the planet must be inside of $a_{pl}<2.2$AU if it is to scatter such particles in as far as an exozodi at $r_{zodi}=1$AU. 

This analysis can also be applied to the inner planet of a chain of multiple planets, as considered later in this work (Sec.~\ref{sec:multi}). In this case, the particle's inclinations and eccentricity will have been excited by the exterior planets before it reaches the inner planet. For example all particles with $e_i\sim 0.2$ by the time they reach the inner planet, can be scattered as far as 1AU by a planet at 5AU. 

%%%%%%%%%%%%%%%%%%%%%%%%%%%%%%%%%%

%%%%%%%%%%%%%%%%%%%%%%%%%%%%%%%%%%

\subsection{Dependence on the planet mass}
\label{sec:mpl}
%%%%%%%%%%%%%%%%%%%%%%%%%%%%%%%%%%

\begin{figure}
\includegraphics[width=0.48\textwidth]{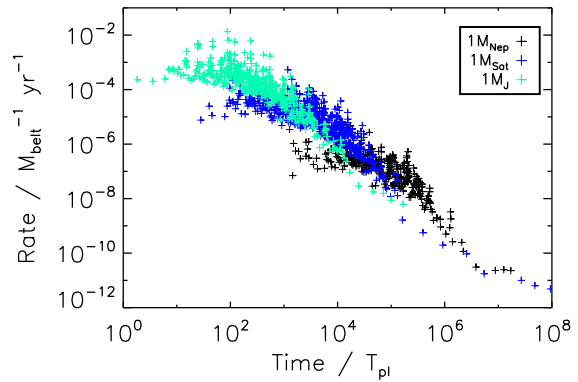}
\caption{Comparing the rate at which particles are scattered inwards by Neptune, Saturn and Jupiter-mass planets on circular orbits at 2AU. The details of the simulations can be found in Table~\ref{tab:onepl}. Higher mass planets scatter more material inwards, but lower mass planets continue to scatter material inwards on longer timescales.}
\label{fig:onepl_mpl}
\end{figure}

%%%%%%%%%%%%%%%%%%%%%%%%%%%%%%%%%%

Given the analytical limit on the ability of a planet outside of 2.2AU on a circular, planar orbit to scatter particles inside of $r_{zodi}=1$AU, we consider the fiducial example of different mass planets on circular orbits at $a_{pl}$=2AU. The simulations are set-up in the manner described in Sec.~\ref{sec:setup} and the rate at which material is scattered inside of $r_{zodi}=\frac{a_{pl}}{2}=1$AU is tracked, as a function of the total mass of a belt spanning between $r_{belt,in}=a_{pl} + \delta a_{chaos}$ and $r_{belt,out}=\frac{3 a_{pl}}{2}= 3$AU. Each cross on the plot in Fig.~\ref{fig:onepl_mpl} represents an individual test particle in our simulations, at the time at which it reached $r_{zodi}$.

An analytical understanding of the dependence of scattering process on planet mass can be found by considering the variance in the size of the chaotic zone (Eq.~\ref{eq:chaos}) with planet mass. Thus, an increase in the total amount of material scattered with planet mass is anticipated. Generic arguments tell us that higher mass planets scatter on shorter timescales. Thus, a significantly increased rate of scattering, at least at early times, is anticipated. This can be seen in Fig.~\ref{fig:onepl_mpl}. Although the high mass planets are good at scattering particles inwards at high rates on short timescales, on longer timescales, low mass planets are better suited to scattering particles inwards at lower rates.

\begin{figure}
\includegraphics[width=0.48\textwidth]{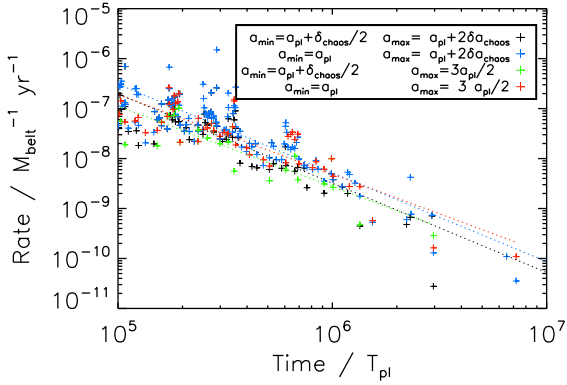}
\caption{The change in the scattering rates with the width of the simulated belt, \ie $a_{min}$ and $a_{max}$, for a $1$M$_{Nep}$ planet at 2AU, as described in Table~\ref{tab:onepl}. The crosses represent individual scattering events, whilst the dotted lines are straight line fits to the data. There is a negligible difference between scattering rates when $a_{max}$ is increased (green and black points), however, a decrease in $a_{min}$ causes a slight increase in the scattering rates, even at late times (red or blue crosses compared to black or green).   }
\label{fig:changeamax}
\end{figure}

%%%%%%%%%%%%%%%%%%%%%%%%%%%%%%%%%%

%%%%%%%%%%%%%%%%%%%%%%%%%%%%%%%%%%

\begin{figure}
\includegraphics[width=0.48\textwidth]{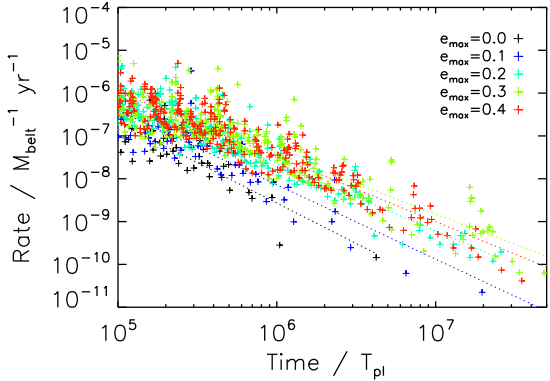}
\caption{The scattering rates of particles that started with different initial eccentricities and inclinations. The simulations are all for a single Neptune mass planet at 2AU and are detailed in Table~\ref{tab:onepl}. The crosses represent individual scattering events, whilst the dotted lines are straight line fits to the data. There is an increase in scattering rates with the particle's initial eccentricity or inclination. }
\label{fig:changeemax}
\end{figure}

%%%%%%%%%%%%%%%%%%%%%%%%%%%%%%%%%%

\subsection{Dependence on the initial conditions of the simulation}
\label{sec:initialconditions}

The precise outcome of N-body simulations depends on the exact initial conditions used. In our case, the initial orbital parameters of the planetesimals in the outer belt can affect the ability of planets to scatter them inwards. Ideally the initial conditions for our simulations would be those found at the end of planet formation process. It is, however, not fully clear what these are. There are many uncertainties in our understanding of the planet formation process and its outcomes, and it is not the purpose of this work to investigate these. We therefore choose to consider an idealised scenario in which the planets are fully formed at the start of our simulations, and the planetesimal belt is situated outside of their orbits.

As planets form they accrete material from their direct surroundings. Thus, a deficit of planetesimals very close to the planet would be expected at the end of planet formation. However, we do not anticipate that the planet fully clears its chaotic zone, given that our initial simulations (Fig.~\ref{fig:1mnep}) showed that some particles in the outer regions of the chaotic zone were only scattered after $10,000T_{pl}$, \ie longer than expected timescales for planet formation. As discussed in Sec.~\ref{sec:setup}, we therefore did not include any particles between $a_{pl}$ and $a_{min} =  a_{pl} + \frac{\delta a_{chaos}}{2}$ in our simulations. Fig.~\ref{fig:changeamax} shows the the range of change that could be anticipated in our scattering rates, if we have either under or over estimated $a_{min}$. At the late times of interest, even the maximum possible change is less than an order of magnitude. 

%Of significance importance for our simulations is the state of the planetesimal belt at the end of planet formation. In our idealised scenario the planetesimal belt is unaffected by planet formation, except potentially that all planetesimals inside of a limiting radius $a_{min}$ are removed. We arbitrarily place this radius at $a_{min}= a_{pl} + \frac{\delta a_{chaos}}{2}$ and consider that deviations from this value will be less relevant at late times. Thus, the early stages of our simulations should not be taken too seriously in understanding our results. Some of the effects of this assumption can be seen in Fig.~\ref{fig:changeamax}. If $a_{min}=a_{pl}$, rather than $a_{min}= a_{pl} + \frac{\delta a_{chaos}}{2}$ (\ie compare the blue, red crosses to the green, black ones), there is a slight increase in the scattering rates, even at late times. However, this change is small compared to the other variations in the scattering rate, for example with planet mass, or even particle initial eccentricity. 

%  where the by comparing either the blue or the black crosses, simulations that used this assumption, with a simulation where the planetesimal belt initially extended right down to the planet (the red crosses) or only started at $a_{min}=??$. There is nothing to suggest that either of the latter two simulations are more valid than our choice, however, this plot shows that at late times ($\sim Myr$) the variation is (much?) less than an order of magnitude. 

In order to reduce the necessary computation time, we also limited the initial semi-major axis range of test particles to less than $a_{max}=a_{pl} + 2\delta a_{chaos}$. Fig.~\ref{fig:changeamax} shows that this does not alter the results more than the statistical variation between individual simulations. For multi-planet systems, there is the potential that three-body resonances or secular terms may affect particles much further out in the disc, however, test simulations found this not to be significant (Sec.~\ref{sec:discussion}).

Arguably the most significant initial condition for our simulations is the initial excitation of the disc. Planet formation models suggest that planetesimals should initially have very low eccentricities and inclinations, damped by the gas disc. For this reason we choose to restrict our initial eccentricities and inclinations to $e_{max}=0.1$ and $I_{max}=10^\circ$ in this work and allow any excitation above these values to occur naturally during the course of the simulations by interactions with the planet(s). This is a already a relatively high limit compared to many observed debris discs \citep{Thebault07}. However, this ignores any potential excitation of the eccentricities and inclinations of the particles during terrestrial planet formation, after the gas disc has dissipated. A different choice of $e_{max}$ and $I_{max}$ could significantly alter our results. Fig.~\ref{fig:changeemax} shows that the rate at which material is scattered inside of $r_{zodi}$ increases with $e_{max}$ and $I_{max}$. To calculate these results it was necessary to start the simulations with particles spread initially over the full belt, so as not the bias the results, given that there is an increase in the size of the chaotic zone with both the initial eccentricity and inclination of the particles \citep{mustill_ecc_chaos, Veras04} not accounted for in the formula of \cite{Wisdom1980}. This increase in the size of the chaotic zone, is responsible for an increase in the number of particles scattered and thus an increase in the rate at which particles are scattered inwards.

%%%%%%%%%%%%%%%%%%%%%%%%%%%%%%%%%%
%% A TABLE WITH ALL OF THE DATA FITTED! and DETAILS OF RUNS
%\begin{minipage}[c]{2\textwidth}
\begin{table}
\begin{center}

\begin{tabular}{l l l l l l }%c c  c |c c }          
       \hline\hline 
                
Label & $a_{min}$ & $a_{max}$ & $e_{max}$ & $I_{max}$ & $N_{pp}$ \\%& $N_{remove}/N_{chaos}$ & $N_{in}/N_{remove}$ &$A$& $\gamma$\ntm \\ % for my reference file name 

\hline
\hline

$1$M$_{Nep}$ & $a_{pl} +\frac{\delta a_{chaos}}{2}$ &$a_{pl}+ 2\delta a_{chaos}$  & 0.1 & $10^\circ$ & 714 \\% & 0.78 &0.39&170$\pm9$ &-1.79$\pm0.26$ \\ % part of 2au_repeat_extra and 2au_repeat stored in fiducial.xdr!! 

$1$M$_{sat}$ &$ a_{pl} +\frac{\delta a_{chaos}}{2}$ & $a_{pl}+ 2\delta a_{chaos}$  & 0.1 & $10^\circ$ & 1000 \\%& 1.15 & 0.66&$13 \pm 3$& $-1.66\pm 0.11$   \\%$1.12\pm 0.49$ \\ % 1msat/2au
$1$M$_J$ &$ a_{pl} +\frac{\delta a_{chaos}}{2}$ & $a_{pl}+ 2\delta a_{chaos}$  & 0.1 & $10^\circ$ & 1000 \\%&  1.15 &0.56&$1\pm 2$&$-1.43 \pm 0.11$  \\%$-0.16 \pm 0.31$ \\ % 1mj/2au

\hline

$e_{max}=0$ & $a_{pl}$ &$\frac{3 a_{pl}}{2}$  & $10^{-4}$ & $0^\circ$ & 1000\\% & $0.67$\ntm & 0.52 &$570\pm40$ &$-1.89 \pm 0.63$\\ % 2au_emax_0
$e_{max}=0.2$ & $a_{pl}$ &$\frac{3 a_{pl}}{2}$  & 0.2 & $20^\circ$ & 1000 \\%& $1.23^2 $&0.82&$66\pm 5$&$-1.58 \pm 0.14$ \\ % 2au_emax_0.2
$e_{max}=0.3$ & $a_{pl}$ &$\frac{3 a_{pl}}{2}$  & 0.3 & $30^\circ$ & 1000 \\%& $1.43^2$ &0.77&$11\pm5$&$-1.41 \pm 0.15$ \\ % 2au_emax_0.4
$e_{max}=0.4$ & $a_{pl}$ &$\frac{3 a_{pl}}{2}$  & 0.4 & $40^\circ$ & 1000\\% & $1.44^2$ &0.76&$42\pm 5$&$-1.52 \pm 0.12$ \\ % 2au_emax_0.4

\hline
$a_{min}=a_{pl}$ & $a_{pl}$ &$a_{pl}+ 2\delta a_{chaos}$  & 0.1 & $10^\circ$ & 949 \\
$a_{max}=\frac{3a_{pl}}{2}$&$a_{pl} +\frac{\delta a_{chaos}}{2}$  &$\frac{3 a_{pl}}{2}$  & 0.1 & $10^\circ$ & 878 \\

Full & $a_{pl}$ &$\frac{3 a_{pl}}{2}$  & 0.1 & $10^\circ$ & 1000 \\

%extra\_1 & 2AU & 3AU & 0.1 & $10^\circ$ & 1000 \\%&1.07&0.37& 17 $\pm$ 5 &$-1.59\pm0.13$ \\ % 2au_repeat_extra
%% add lots of simulations together! 

%lots  & 2AU & 3AU & 0.1 & $10^\circ$ & 4000 &1.05&0.47 &$869\pm 9$& $-1.53\pm 0.16$ \\ %  2au_repeat + 23  + 2au_repeat + 2au_extra

%repeat   & 2&3 & 0.1 & $10^\circ$ & 1000 &0.51&1.17& $171\pm 43$& $-1.74 \pm 0.29$  \\ % 2au_repeat
%lots  & 2 & 3 & 0.1 & $10^\circ$ & 2000 & 1.15&0.52& $-1.84\pm 0.34$ \\ %  2au_repeat + 23
%extra\_2 & 2AU & 5AU & 0.1 & $10^\circ$ & 1000 \\%& 1.05 &0.47& $12\pm 8  $ & $-1.53\pm 0.16  $ \\ % 2au_repeat_extra_extra

%%%%% SHOULD THE NREMOVE/NCHAOS for mpl different be for a belt from delta_a to 3 AU or 2-3AU 

%% what about C=1.3 or C=2.0- BE CAREFUL! so far C=2.0 used.. 
 
\hline\hline   
\end{tabular}

\end{center}

\caption{Details and results of the various one planet simulations. All simulations are for a $1M_{Nep}$ mass planet, unless otherwise stated. } 

\label{tab:onepl}

\end{table}

\section{Multi-planet Systems}%: chains of equal mass planets separated by $\Delta R_H$}
\label{sec:multi}

A more realistic planetary system is one in which there are multiple planets and the outer planetesimal belt is significantly further from the star. There is good evidence for the prevalence of multi-planet systems with a wealth of different architectures\footnote{162/580 multi-planet systems according to exoplanets.org on 18 June 2012}. The full parameter space is clearly  too large to investigate in this work, even if we restrict ourselves to $N_{pl}$ planets on circular orbits, in the same plane, interior to a planetesimal belt. For simplicity, we therefore focus on the example of a chain of equal mass planets, separated equally in units of their mutual Hill´s radii. This system should be considered a representative system, from which it is possible to better understand the scattering processes and the way that they apply to other systems. 

The simulations are set-up in the same manner as described in Sec.~\ref{sec:simulations}. We place the outer belt at $r_{belt,in}=62$AU, to match with the observations of the outer belt in the Vega system \citep{muller}, although we anticipate that our simulations can be scaled to a limited range of different belt radii. In the same manner as for the single planet system, the outer planet is placed such that it sculpts the inner edge of the outer belt, at $a_{pl,out}= r_{belt,in}-\delta a_{chaos}$. Equal mass planets are then placed interior to this, separated by $\Delta  R_H$. In order to ensure that the planetary systems considered actually scatter material inside of $r_{zodi}=1$AU, we consider the conservation of the Tisserand parameter, and the limit this provides on how far particles can be scattered by a planet (Eq.~\ref{eq:qmin}). This time, we anticipate that the scattering by multiple outer planets excites the particles eccentricities and inclinations before they are scattered by the interior planet. Therefore, we place a limit on the interior planet such that all particles with $e>0.2$ can be scattered inside of $r_{zodi}$=1AU. This implies that the planet should be at 5AU. Therefore we `fill' our planetary  system with planets equally spaced in terms of their Hill's radii down to $a_{fill, in}$=5AU. The number of planets required to `fill' the planetary system ranges between 3 and 10 (as can be seen in Table~\ref{tab:multi}). A choice of a larger value for $a_{fill,in}$ could reduce the number of particles scattered inside of $r_{zodi}$ to zero. It is therefore worth noting that our simulations are more representative of the maximum, than minimum scattering rates. % Natually, this may well occur in some planetary systems, although it is not necessary to use N-body simulations to simulate this. It is, however, worth remembering that our simulations focus on systems in which the scattering rates are non-zero. }

%The planetary systems produced and investigated here are, therefore, very contrived and specifically designed to scatter material inwards to the location of the exozodi. The intention is not to imply that it is likely that any real planetary system should possess such a contrived architecture. In particular, since it is unlikely that planet formation is equally efficient and thus able to produce equal mass planets throughout a planetary system. Instead, the planetary systems investigated here should be considered as examples from which it is possible to understand the scattering processes and estimate the scattering rates in any planetary system. The application of these results to different planetary systems is discussed further in Sec.~\ref{sec:scaling}. 

%}

\begin{figure}
\includegraphics[width=0.48\textwidth]{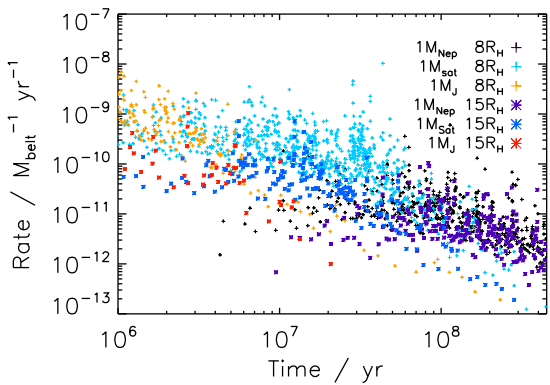}
\caption{The rate at which particles are scattered, by a chain of equal mass planets of $1M_{Nep}$ (dark purple) $1M_{Sat}$ (blue) or $1M_J$ (red/orange), spaced equally in units of their mutual Hill's radii by either 8 (crosses) or 15 (stars) $R_H$. The rate is plotted as a fraction of the outer belt mass, in a belt spanning from 62AU to 93AU, and particles are scattered inside of $r_{zodi}=1$AU, as described in Sec.~~\ref{sec:multi} and Table~\ref{tab:multi}. These results can be scaled to different systems, as described in Sec.~~\ref{sec:scaling}. The scattering rate is higher on shorter timescales with higher mass planets that are closely spaced, whilst on longer timescales tightly packed low mass planets scatter at a higher rate.  }
\label{fig:multi}
\end{figure}

Using the same procedure as for the single planet case (described in Sec.~~\ref{sec:onepl}), the rate at which particles are scattered inside of $r_{zodi}=1$AU of the star in our N-body simulations is calculated. Again, each particle is assigned a fraction of the outer belt mass and the rate at which material is scattered inwards is calculated by considering the time between adjacent scattering events. Again, we arbitrarily consider outer belts that extend to $r_{belt,out}= \frac{3}{2} r_{belt,in}= 93$AU. We vary the planet masses, $M_{pl}$ and separation $\Delta R_H$ between our simulations. The dependence on the stellar mass $M_*$ and the position of the outer belt, $r_{belt,in}$ should also be noted (discussed further in Sec.~\ref{sec:discussion}). The results are shown in Fig.~\ref{fig:multi}, for the different simulations listed in Table~\ref{tab:multi}.

Fig.~\ref{fig:multi} clearly shows that all simulations follow the same general trend; the scattering rates decrease with time. Above this, there are two clear dependencies on the architectures of the planetary system, one on the planet mass and one on their separation. Firstly, as discussed in Sec.~\ref{sec:mpl}, we expect higher mass planets to scatter more material on shorter timescales. This is clearly seen in Fig.~\ref{fig:multi}, where the scattering rates increase with planet mass between the purple blue crosses and yellow crosses. Secondly, the more widely spaced the planetary system, the lower the scattering rates \eg the decrease in scattering rates between the yellow crosses ($8R_H$) and red crosses ($15R_H$). This follows simply from the fact that the wider the separation of the planets, the smaller the proportion of the particles scattered by one planet that are on orbits that interact with the next inner planet, and so forth for each pair of planets. The planet mass clearly dominates the scattering rates.

%To summarise, the important parameters that can be varied in these simulations are the planets´ masses, $M_{pl}$, the stellar mass $M_*$ (or their ratio $\mu=\frac{M_{pl}}{M_*}$), the position of the outer belt, $r_{belt,in}$, the separation of the planets, $\Delta$, $r_{zodi}$ and $a_{fill}$. We will later show that the dependence on the latter two parameters is less critical (within a reasonable range). We vary the planets' masses and separation and use our simulations to discuss the effects of varying the other parameters. 

%%%%%%%%%%%%%%%%%%%%%%%%%%%%%%%%%%

%%%%%%%%%%%%%%%%%%%%%%%%%%%%%%%%%%

\begin{table*}
\begin{tabular}{ c c c c c c c c c c c}
\hline \hline
Label &&Stellar & \multirow{2}{*}{$r_{belt,in}$} & \multirow{2}{*}{ $r_{belt, out}$ }&No. of  & Planet & Planet & Planet semi-major axes  & \multirow{2}{*}{ $r_{zodi}$} & \multirow{2}{*}{ $a_{fill}$ }\\
&&Mass &  & & planets &Mass & Separation& & & \\
&&$M_\oplus$&AU &AU & & $M_J$& $R_H$ & AU & AU & AU \\ \hline \hline

% solar mass
% 10rh
 \multirow{6}{*}{Original}& 1& 1&62 & 93  & 4 &1 & 8
 & 51.0,	25.9,	13.2,	6.7	& 1 & 5 \\
&2 & 1&  62 & 93 & 6	& 0.3&8 &53.8,	34.5,	22.2,	14.2,	9.1,	5.9		& 1 & 5 \\
&3 & 1  &  62 & 93 & 9 & 0.053& 8	&56.7	44.3	34.6	27	16.5	12.9	10.1	7.9	6.2	& 1 & 5 \\
% 20- rh 
& 4& 1& 62 &93 & 2 &1 & 15 &51.0,	10.8	& 1 & 5 \\
& 5& 1& 62 & 93 & 3	& 0.3&15 &53.8,	21.1,	8.3	& 1 & 5 \\
&6& 1&  62 & 93 & 5& 0.053& 15	&56.7,	34.3,	20.8,	12.6,	8	& 1 & 5 \\ \hline \hline

%% NB technically delta_chaos was wrong for these.. hmm.. shouldn't make a big difference, i hope! 

% VEGA
%% SHOULD I SCALE FOR A DIFFEREN RBELT OUT FOR VEGA AND TAU CETI??
 \multirow{6}{*}{Vega}&1v& 2.16 &62 & 120 & 4 &2.16 & 8 & 51.0	25.9	13.2	6.7	& 1 & 5 \\
&2v&2.16& 62 & 120& 6	& 0.65&8 &53.8	34.5	22.2	14.2	9.1	5.9		& 1 & 5 \\
&3v&2.16 & 62 & 120 & 9 & 0.11&8	&56.7	44.3	34.6	27	16.5	12.9	10.1	7.9	6.2	& 1 & 5 \\  

& 4v& 2.16& 62 &120 & 2 &2.16 & 15 &51.0,	10.8	& 1 & 5 \\
& 5v& 2.16& 62 &  120 & 3	& 0.65&15 &53.8,	21.1,	8.3	& 1 & 5 \\
&6v& 2.16& 62 &  120 & 5& 0.11& 15	&56.7,	34.3,	20.8,	12.6,	8	& 1 & 5 \\
\hline

\multirow{6}{*}{$\eta$ Corvi}&1e&   1.42 &100 &150 & 4 & 1.42 &6.5 &82.3,	41.8,	21.3,	10.8 &   1.6 & 8 \\
&2e& 1.42 &100 &150  & 6& 0.43&6.5 &86.8,	55.6,	35.8,	22.9	,14.7,	9.5   & 1.6 & 8 \\
&3e& 1.42 &100 &150  & 9& 0.08& 6.5  &91.5,	71.5,	55.8,	43.5	,26.6,	20.8,	16.3,	12.7 & 1.6 & 8 \\

&4e& 1.42 &100 &150 &  2& 1.42 & 12& 82.3,	17.4  & 1.6 & 8 \\
&5e& 1.42 &100 &150 &3 & 0.43 & 12  &86.8, 	34.0, 	13.4 & 1.6 & 8 \\
&6e& 1.42 &100 &150  &5&  0.08& 12  &91.5,	55.3,	33.5,	20.3,	12.3 & 1.6 & 8 \\

%1&$\tau$ Ceti  & 0.78 &10 &55  & 4 &0.78 & 8 & 	8.5,4.3,	2.2,	1.1	& 0.16 & 0.8 \\
%2 &$\tau$ Ceti  & 0.78 &10 & 55& 6	& 0.23&8 &9.0, 5.8,	3.7,	2.4,	1.5,	1.0	& 0.16 & 0.8 \\
%3 &$\tau$ Ceti  & 0.78 &10 & 55& 9 & 0.04&8	&9.5, 7.4,	5.8,	4.5,	2.8,	2.2,	1.7,	1.3,	1.0	& 0.16 & 0.8 \\
 
%4& $\tau$ Ceti &0.78& 10 & 55& 2 &0.78 & 15 &8.5,	1.8& 0.16 & 0.8 \\
% 5&$\tau$ Ceti  & 0.78& 10 &55  & 3	& 0.23&15 &9.0,	3.5,	1.4& 0.16 & 0.8 \\	
%6&$\tau$ Ceti  & 0.78& 10& 55 & 5& 0.04& 15	&9.5 ,	5.7 ,	3.5,	2.1,	1.2	& 0.16 & 0.8 \\

		\hline\hline																	
	\end{tabular}
\label{tab:multi}
\caption{A summary of the parameters for the chains of equal mass planets used in the simulations of Sec.~~\ref{sec:multi}, shown in Fig.~\ref{fig:multi}. The bottom half of the table describes scaled versions of these simulations that could be applied to Vega or $\eta$ Corvi.} %. $r_{zodi}=1$AU, $a_{fill}=5$AU for Vega and $r_{zodi}=0.16$AU, $a_{fill}=0.8$AU for $\tau$ Ceti.}																		
\end{table*}

%%%%%%%%%%%%%%%%%%%%%%%%%%%%%%%%%%

%%%%%%%%%%%%%%%%%%%%%%%%%%%%%%%%%%

%\begin{figure}
%\includegraphics[width=0.48\textwidth]{new_rate_multi_afillin.eps}
%\caption{Scattering rates from a chain of Jupiter mass planets separated by $8R_H$, but varying the number of planets, or the minimum limit on the planet's semi-major axis, $a_{fill}$. $a_{fill}=5$AU is our fiducial simulation with 4 planets, $a_{fill}=10$AU has only 3 planets and we found that for $a_{fill}>15$AU no particles are scattered inside of $r_{zodi}=1$AU. This implies that the scattered particles only gain eccentricities up to $e\sim 0.3$, from the conservation of the Tisserand parameter and the limit on how far particles can be scattered (Eq.~\ref{eq:qmin}.}
%\label{fig:afillin}
%\end{figure}

%%%%%%%%%%%%%%%%%%%%%%%%%%%%%%%%%%

\subsection{Application of these results to any system}
\label{sec:scaling}

The simulations presented in this section are for very specific planetary systems, with strict constraints on its architecture. This was necessary in order for the current investigation to be feasible. Clearly, these do not mimic many, if any, real planetary systems. In fact, the current observations of multi-planet systems find a huge variety of planet masses and separations. Nonetheless, the results from these simulations can be used to make useful conclusions about the scattering rates in more realistic planetary systems. Firstly, having understood the broad dependences of the scattering rates on planet separation and mass, we are able to conclude that the highest scattering rates, at late times, occur for tightly packed, low-mass planets. Thus, our simulations with chains of Neptune mass planets separated by $8R_H$ are representative of some of the highest possible scattering rates after millions of years of evolution, beaten only by chains of lower mass planets, packed even closer to the stability limit. We therefore consider that our simulations are broadly representative of an approximate upper limit on the scattering rates possible in any planetary system, although, it is worth noting, as will be discussed in Sec.~\ref{sec:discussion}, that we may have missed a few systems with very specific architectures that enhance the rates still further. Secondly, the results presented here can be scaled, such that the scattering rates in a system of a known architecture can be estimated. In this manner the scattering rates in a real planetary system, with planets on known orbits, could be estimated.

 In this work, we investigated the change in the scattering rates between chains of equal mass planets, evenly spaced in units of their Hill's radii ($\Delta R_H$), with different planet masses ($M_{pl}$) and separations ($\Delta$). To do this we fixed the stellar mass ($M_*$), the radial location of the outer belt, ($r_{belt,in}$, $r_{belt,out}$), the radial location of the exozodi ($r_{zodi}$) and a parameter that we term $a_{fill}$, that specifies the number of planets, by providing an inner radial limit on their possible orbits. Real planetary systems are unlikely to have the values used for these parameters. We therefore discuss the effects of changing the different parameters on the scattering rate, in such a way that we can understand and anticipate the behaviour in any planetary system.

Firstly considering planetary systems of the form described in this section, with equally separated, equal mass planets, the scattering rates that we calculated (\ie shown in Fig.~\ref{fig:multi}), can be applied almost exactly to another system in which the orbital timescales of the planets/planetesimals are different. In other words, systems where the radial location of the outer belt or the stellar mass is changed. In this manner the results of our simulations can be applied to planetary systems with outer planetesimal belts at any location, with the assumption that the radial location of the exozodi scales with the radial location of the outer belt. If the radial location of the outer belt is scaled by a factor, $f$, then the same scattering rates can be applied to the system by scaling the time axis by a factor of $a_{pl,out}^{3/2} \sim r_{belt,in}^{3/2}$. Similarly, for stars of different stellar mass, there is a small change in the scattering timescales that can be found by scaling the time axis by a factor $M_*^{1/2}$.

There are three other parameters that can affect the scattering rates; the width of the outer belt ($r_{belt,out}$), the location of the exozodi ($r_{zodi}$) and the innermost planet's semi-major axis, $a_{fill}$. The width of the outer belt is in essence unimportant, as long as it is greater than the planet's chaotic zone. It will however change the values for both the total outer belt mass, and the fraction of it, that is scattered inwards, in a manner that can be readily calculated using Eq.~\ref{eq:mpart}. 

The location of the exozodi ($r_{zodi}$) is crucial. Scattering rates generally scale with $r_{zodi}$, unless $r_{zodi}$ is too small that no (or very few) particles can be scattered sufficiently far inwards (see discussion of the analytical limit in Sec.~\ref{sec:tiss}). Changing the location of the exozodi produces differences in the scattering rate of the same order of magnitude as the difference caused by changing the planet separation ($\Delta$).

 On the other hand, the value of the parameter $a_{fill}$ used here, was chosen to ensure that particles are scattered to the location of the exozodi, $r_{zodi}$. In a real planetary system, the innermost planet could take any position, and the scattering rates will depend critically on this position. Test simulations find that scattering rates do not vary significantly with $a_{fill}$, or equivalently the position of the innermost planet, so long as the planet separation does not increase. This only applies if $a_{fill}$ is not decreased so much that the inner planet is within a Hill's radius of the exozodi, or increased so much that particles can no longer be scattered from $a_{fill}$ to $r_{zodi}$. For the example system used in this work with $a_{belt,out}=62 r_{zodi}$, analysis of our simulations found that particles are generally excited up to eccentricities of around $e\sim0.3$ before being scattering by the planet. Thus, using Fig.~\ref{fig:tiss_limit} we anticipate that increasing $a_{fill}$ up to 10 AU, does not significantly alter the scattering rates. We verified this with test simulations. In other words, taking the example of the chain of 4 Jupiter mass planets separated by $8R_H$, the innermost planet is not required in order to retain the scattering rates at the calculated values. However, if $a_{fill}$ were to be increased beyond 15AU, no particles would be scattered inwards. It is therefore easy to design a planetary system in which the scattering rates tend to zero, by applying Fig.\ref{fig:tiss_limit} to that system.

% changing the value of the parameter, $a_{fill}$, does not significantly affect the results, unless it is increased such that particles can no longer be scattered from $a_{fill}$ to $r_{zodi}$. This limit can be determined approximately by considering the minimum pericentre for scattered particles with  analysis}. T

Now, we consider the broad outcomes of these results and their applicability to any planetary system. In general, the results presented in this work only further our understanding of planetary systems with a single, outer planetesimal belt and planets on circular, co-planar orbits, interior to the belt. Let us first consider chains of planets of non-equal mass. The outer planet controls the amount, and the rate, at which material is scattered into the inner planetary system. Thus, the outer planet dominates the scattering rates, whilst the inner planets modify these rates slightly by controlling how long it takes for particles to be passed inwards to the exozodi. This timescale depends on the mass of the planets. The best way to illustrate this is to consider the simple example of a chain of Neptune mass planets, that ordinarily scatter a small amount of material inside of $r_{zodi}$ on long timescales (see Fig.~\ref{fig:multi}). If all but the outer planet were replaced with higher mass (\eg $\sim 1M_J$) planets, whilst retaining the same separation in Hill's radii, then material would continue to be scattered out of the outer belt on the same timescales, but would be passed from the outer belt to $r_{zodi}$ more swiftly. Therefore, the rate of scattering of material inside of $r_{zodi}$ would fall off more steeply with time. If the interior planets were of lower mass than the outer planets, they could do the inverse and cause a bottleneck, delaying the scattering process.

Now, all that remains in order to be able to apply our results to any planetary system, is a consideration of the effect of changing the separation of the planets. Although our simulations clearly show that more widely spaced systems scatter at somewhat lower rates, as discussed in Sec.~\ref{sec:multi}, the effect on the scattering rates is not as significant as that caused by changing the planet mass. The separation of the planets can, however, be increased so much, that a gap in the planetary system is produced, over which very few particles are scattered. The separation at which this occurs can be estimated by considering when the planets are so widely separated that $q_{min} (T_i) < a_{pl} - \Delta R_H $, or referring to \cite{bonsor_tiss} for a more detailed calculation. To take an example, for a chain of Jupiter mass planets, if any pair of planets were separated by more than $20R_H$, very few particles could be passed between them.

%For a complicated chain of multiple planets, this analysis could be used to estimate the timescales on which particles are passed between the different planets. However, compared to changing the planet mass, the effect of changing the planet separation is not as significant, unless, the planets are so widely separated that no particles can be scattered between them. A detailed description of how to calculate the limit where there is a 'blockage' in a chain of multiple planets was described in detail in \cite{bonsor_tiss}, based on the initial value and conservation of the Tisserand parameter. A simple estimate can be made by considering Eq.~\ref{eq:qmin}, no further scattering occurs if the planets are so widely separated that $q_{min} (T_i) < a_{pl} - \Delta R_H $. Therefore, the most important consequence of changing the separation of the planets can be the production of a block in the scattering chain. Apart from this a change in the separation of the planets only makes small changes to the scattering rates, compared to the changes made by changing the planets' masses.  

To summarise, the arguments presented here can be used to calculate exactly the scattering rates in planetary systems that follow the mould of our example and to estimate the broad behaviour in any other planetary system.

\subsection{The collisional evolution of  the outer planetesimal belt}
\label{sec:alim_multi}

The outer planetesimal belt is a collisionally evolving system, like any debris disc. We have not taken the collisional evolution of the disc mass into account in our calculation of the rate at which material is scattered inwards, shown in, for example, Fig.~\ref{fig:multi}. Each individual data point in this plot was calculated by assigning each test particle a fraction of the total outer belt mass at the instant it was scattered inwards. Of course, if the disc is collisionally evolving, the total outer belt mass could potentially be different for each particle. Taking this into account, the new version of Fig.~\ref{fig:multi}, would have an even steeper fall off with time. For practical purposes, the data in Fig.~\ref{fig:multi} will generally be used to assess the total outer belt mass required in order to obtain an observed scattering rate. This should be close to the disc's current mass, because in the general the timescales for the collisional evolution are not significantly faster than the scattering timescales.

If we make the assumption that material must be scattered inwards at a rate of at least $10^{-9}M_\oplus/$yr to resupply an exozodi, the simulations shown in Fig.~\ref{fig:multi} can be used to determine the minimum mass required an the outer belt at 62AU. It is important to consider whether or not such a mass can survive against collisions, at the given radial location, for the age of the system. This can be done analytically, for example using the models of \cite{wyatt07hot} and Eq.~\ref{eq:mmax}. In fact, given that the results of our simulations can be readily scaled to systems with outer belts at different radial locations (as discussed in Sec.~\ref{sec:scaling}), the minimum mass required can be determined for belts at a variety of different radii. This scaling necessarily assumes that the location of the exozodi scales with the outer belt. The best way to illustrate this is to take the scattering rates, $R_{s}$, calculated and shown in Fig.~\ref{fig:multi}. The disc mass required to produce these scattering rates, by an outer belt at any (reasonable) radial location, $r_{belt,in}$ (\ie scaling the rates of Fig.~\ref{fig:multi} with the orbital timescales), is given by 
\begin{equation}
M_{disc} = \left (\frac{ R_s / M_\oplus yr^{-1} }{10^{-9}/ M_\oplus yr^{-1}} \right ) \left (\frac{ r_{belt,in / AU}}{62 AU} \right)^{3/2}   \; \; M_\oplus.
\label{eq:scale}
\end{equation}

\begin{figure}

\includegraphics[width=0.48\textwidth]{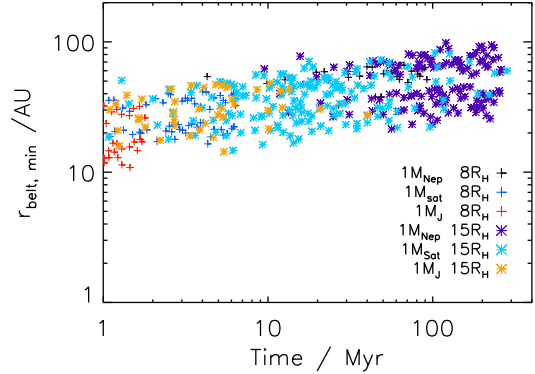}

\caption{The minimum radius of an outer belt, $r_{belt,min}$, in which sufficient mass to scatter material inwards at a rate of $10^{-9}M_{\oplus}/$yr can survive against collisions, as a function of time. Each cross is taken from the simulations shown in Fig.~\ref{fig:multi}, scaled to different radii for the outer belt and planets using Eq.~\ref{eq:scale} and converted to a minimum radial location for the belt using Eq.~\ref{eq:alim}. }
\label{fig:alim_multi}
\end{figure}

We now have a requirement on the minimum mass needed in an outer belt, at a range of radial locations, if the outer belt is to supply material scattered inwards at a rate of at least $10^{-9}M_\oplus/$yr. Contrasting this with the collisional evolution of the disc, it is possible to determine which of these discs can survive against collisions for the age of the system. In other words a minimum limit on the radius required for the outer belt can be placed. This limit is shown in Fig.~\ref{fig:alim_multi}. Given the assumptions, this means that an outer belt that resupplies an exozodi must lie beyond the radial limit shown in Fig.~\ref{fig:alim_multi}. Each cross in this plot is taken from the simulations shown in Fig.~\ref{fig:multi}, where the minimum mass required in the disc is calculated for a range of radial locations using Eq.~\ref{eq:scale} and the minimum possible radial location determined by substituting these masses into Eq.~\ref{eq:rmin}, to obtain:

\begin{eqnarray}
&&r_{belt,min} =K^{26/17} \left (\frac{t^2 (R_s / M_\oplus yr^{-1})^2}{(10^{-9}/ M_\oplus yr^{-1})^2 \; (62 \mathrm{AU})^3} \right ) ^{3/17}  \mathrm{AU} \\ \nonumber
&&K=\frac{6.8 \times 10^2 (M_*/M_{odot})^{4/13} \langle e \rangle^{5/13}}{(D_{max}/km)^{3/13} (\rho/kgm^{-3})^{3/13} (\frac{dr}{r})^{3/13} (Q_D^*/Jkg^{-1})^{5/26}} \\\nonumber
\label{eq:alim}
\end{eqnarray}

This makes the assumption that the belt width is small and takes the values for the parameters listed in Sec.~\ref{sec:mass}. Fig.~\ref{fig:alim_multi} clearly shows that because relatively massive planetesimal belts are required to scatter material inwards at the required rates, the belt must be reasonably far from the star in order that such a mass can survive against collisions for the typical ages of exozodi systems. This of course assumes that the collisional evolution of the belt commenced very early in the star's evolution and that all bodies in the disc are involved in this collisional evolution. It also assumes that the belt width is small and fixed at $\frac{dr}{r}=0.5$, that the location of the exozodi scales with the radial location of the outer belt and that a scattering rate of $10^{-9}M_\oplus/$yr is reasonable requirement to retain the observed dust levels in an exozodi. In this case, Fig.~\ref{fig:alim_multi} shows that in order for an outer planetesimal belt to scatter dust inwards at this level for longer than 100Myr, it must be outside of $\sim20 $AU, otherwise insufficient mass can survive against collisions. This is an important conclusion because many observed debris discs are inside of this radius.

%%%%%%%%%%%%%%%%%%%%%%%%%%%%%%%%%%

%%%%%%%%%%%%%%%%%%%%%%%%%%%%%%%%%%

\subsection{The detectability of the outer belt } 
\label{sec:nobelt}
Our simulations can be used to place limits on the required masses and radial locations of an outer planetesimal belt, in order that it could potentially scatter material inwards at sufficiently high rates to retain the currently observed levels of dust in exozodi. These limits can be compared with our ability to detect such discs.

\begin{figure}
\includegraphics[width=0.48\textwidth]{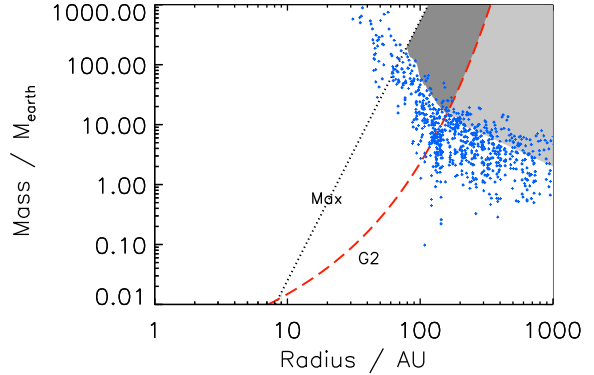}
\caption{ Outer belts capable of scattering material inwards at $10^{-9}M_\oplus/$yr after 100Myr, around a G2 ($1M_\odot$) star, with a chain of $1M_{Sat}$ planets separated by $8R_H$, that can be detected by Spitzer at 70$\mu$m are shown in dark grey, whilst those belts that cannot be detected are shown in light grey. This is the lowest limit of the simulations shown in Fig.~\ref{sec:multi}. The dotted line shows the maximum mass that can be retained against collisions (Eq.~\ref{eq:mmax}) for 100Myr and the dashed line shows the Spitzer detection limit for a G2 star. Full details of this plot are described in Sec.~\ref{sec:nobelt}.}

%A plot of the detection limits on the detection of outer planetesimal belts with Spitzer at 70$\mu$m, based on the radius and mass of the belt, assuming that Spitzer is calibration limited with $\frac{F_{disc}}{F_*}>0.55$ required for detection. Discs can be detected around a G2 ($1M_\odot$) stars if they lie above the blue, red dotted lines, respectively. Sufficient mass can only survive for 100Myr in discs beneath the black dotted lines. Discs above and to the right of the solid coloured lines contain sufficient mass to retain scattering rates of $10^{-9}M_\oplus/$yr after 100Myr. These were calculated using the simulations for chains of equal mass planets in Fig.~\ref{sec:multi}, taking the scattering event closest to 100Myr and scaling the rate calculated from this point by the orbital periods. Thus, the dark grey shaded area shows debris discs capable of scattering at $10^{-9}M_\oplus/$yr after 100Myr and that can be detected, whilst the lighter gray area shows discs capable of the same scattering rates, but undetectable. Further details of this plot are explained in Sec.~\ref{sec:nobelt}.    }
\label{fig:mass_detlim}
\end{figure}

Taking our simulations for chains of multiple equal mass planets, shown in Fig.~\ref{fig:multi}, we calculate the minimum disc mass required to scatter material inwards at a fiducial rate of $10^{-9}M_\oplus/$yr, at an age of 100Myr. This is the total mass, in bodies of all sizes, in a disc between $r_{belt,in}=a_{pl} + \delta a_{chaos}$ and $r_{belt,out}=\frac{3 r_{belt,in}}{2}$. As discussed in Sec.~\ref{sec:scaling} the scattering rates, and thus the minimum required disc mass, can be scaled to outer belts at any radial location, although only with the assumption that the position of the exozodi, $r_{zodi}$ scales in a similar manner. Thus, every point in Fig.~\ref{fig:multi} can be considered to represent the scattering rate after 100Myr in a belt at a different radial location. The radial location and minimum required masses for the outer belts to scatter material inwards at rates higher than $10^{-9}M_\oplus/$yr are shown by the blue crosses in Fig.~\ref{fig:mass_detlim}, for a chain of $1M_{Sat}$ planets separated by $8R_H$. This is the simulation with the largest detectable parameter space of any of our simulations. This plot reiterates the fact that massive, large radii planetesimal belts are the most adept at scattering material inwards. 

Some constraints on the existence of the discs outlined on  Fig.~\ref{fig:mass_detlim} can be made by considering their collisional evolution. As discussed in Sec.~\ref{sec:mass}, only belts below a maximum mass, can survive at a given radial location for a given age. This maximum mass is plotted by the dotted line on Fig.~\ref{fig:mass_detlim}, calculated using Eq.~\ref{eq:mmax}, taking the parameters of \cite{wyatt07}, listed in Sec.~\ref{sec:mass}.

The next question to consider is which of these discs could be detectable. One way to detect a debris disc is by an infrared excess, for example at $70\mu$m. For this plot, we consider Spitzer's detections limits, as Spitzer surveys of debris discs are comprehensive and well understood, despite the better detection capabilities of \Herschel or ALMA (for colder debris discs). Spitzer observations are calibration limited, requiring a ratio of the disc to stellar flux above $\frac{F_{disc}}{F_*}>0.55$ for a detection. By considering a simple model for a debris disc, with a steady-state collisional cascade, $n(D)dD\propto D^{7/2}dD$, for grains between the blow-out size, $D_{bl}$ and  $D_{max}=60$km \citep[e.g.][]{wyattreview}, the minimum detectable disc flux can be turned into a minimum detectable disc mass, $M_{det}$, of: %, the total mass of which is given by \citep{} $ M_{disc}=2.7 f (r /AU)^2 \sqrt{(D_{bl}/\mu m)(D_{max}/km)} M_\oplus$, such that

\begin{eqnarray}
&M_{det}= 1.6 \times 10^{10} R_{\mu,lim}\sqrt{(D_{bl}/\mu m)(D_{max}/km)} \\ \nonumber
& \left (\frac{L_*/L_{\odot}}{ (T_*/K)^4}\right )\frac{B_\nu (\lambda, T_*)}{B_\nu (\lambda, T_{disc})} \; \; M_\oplus \nonumber
\end{eqnarray}
where the $B_\nu$ refers to the Wien function describing black-body emission at a given wavelength, $\lambda$ or frequency $\nu$, at either the stellar temperature, $T_*$ or the disc temperature, $T_{disc}$. Discs that can be detected are those than lie above the dashed line in Fig.~\ref{fig:mass_detlim}, for a G2 ($1M_\oplus$) star. The dark grey shaded area shows all discs that could  be detected by Spitzer and scatter material in at a sufficiently high rate to resupply an exozodi at $10^{-9}M_\oplus/$yr, at an age of 100Myr. The lighter shaded area shows the parameter space in which debris discs capable of scattering material inwards at a sufficiently high rate could lie, but not be detected with Spitzer at $70\mu$m.

The clearest conclusion from this plot is that debris discs exist that could supply scattering at the required rates in systems older than 100Myr. Such debris discs are at large radial distances and massive. It is evident from Fig.~\ref{fig:mass_detlim} that such large radii discs are the population about which least information is known, although this may change in the near future with instruments such as ALMA or SCUBA 2.

This plot should be viewed with several uncertainties and restrictions in mind. Firstly, all the lines on the plot vary with the stellar luminosity or mass of the star. This plot assumes that required scattering rate is $10^{-9}M_\oplus/$yr, a value which is uncertain even for the Vega system and may vary significantly between different systems. It also makes the potentially unrealistic assumption that the radial location of the exozodi scales with that of the outer belt.

%%%%%%%%%%%%%%%%%%%%%%%%%%%%%%%%%%%%%%%%%%%%%%%%%%%%%%%%%%%%%%%%%%%%%%%%%%%%%%%%%%%%%%%%%%%%%%%%%%%%%%%%%%%%%%%%%%%%%%%%%%%%%%%%%%%%%%%%%%%%%%%%%%%%%%%%%%%%%%%%%%%%%%%%%%%%%%%%%%%%%%

%%%%%%%%%%%%%%%%%%%%%%%%%%%%%%%%%%

%%%%%%%%%%%%%%%%%%%%%%%%%%%%%%%%%%

\subsection{Application to observed systems }
The number of systems with detected exozodi is limited, only a handful of the detected systems also have outer planetesimal belts and there are very few planet detections. For these reasons, we decide to focus here on two example systems, both of which have resolved outer belts, and detailed modelling of the inner exozodi. Neither of the systems have any evidence for planetary companions. The two example systems were chosen to contrast. Vega is a 450Myr old A star, whilst $\eta$ Corvi is an older (Gyr), sun-like star, with an extremely large infrared excess. We apply the simulations presented in this work, to these systems, in order to determine the possible scattering rates that could occur, if we make the hypothesis that chains of planets orbit between the inner and outer dust discs. We then compare these scattering rates with observations of the exozodi and outer belt, and assess the feasibility of such scattering to solve the conundrum of the high levels of exozodiacal dust observed.

\subsubsection{Vega}
\label{sec:vega}
Vega's outer belt was the first debris disc to be detected \citep{aumann1984} and has since been observed many times at a variety of different wavelengths. The exact properties derived from observations of the outer belt vary in the literature, but here we consider the best fit model of \cite{muller}, that takes into account observations from the near-infrared through to sub-mm wavelengths. Their best fit model finds a dust belt between 62 and 120 AU, with $5.09M_\oplus$ of dust and a total mass of $46.7M_\oplus$, which collisionally evolved from an initial mass of $55.5M_\oplus$. Fitting of isochrones finds an age of $455\pm13$ Myr and a mass of $2.157 \pm0.017 M_\oplus$ \citep{Yoon2010}. Exozodiacal dust was detected around Vega by \cite{Absil06} and again by \cite{Defrere_Vega}. The best-fit to the observations finds a mass of $10^{-9}M_\oplus$ in small dust grains between 0.2$\mu$m and 1mm in size \citep{Defrere_Vega}. A collisional timescale of $\sim 1$yr is derived. This suggests that if the disc is to be retained at its current levels it needs to be replenished at a rate of at least $10^{-9}M_\oplus/$yr. 

The scattering rates determined in our multi-planet simulations described in Sec.~\ref{sec:multi} and shown in Fig.~\ref{fig:multi} can be applied to Vega. Without any limits on planetary companions less than several Jupiter mass \citep[e.g.][]{Hinkley07, Itoh2006}, we are free to hypothesise the existence of any inner planetary system. If Vega is really orbited by as yet undetected planets, they may or may not resemble the chains of multi-planet systems considered in this work. However, by considering the hypothesis that there are chains of equal mass planets, interior to Vega's outer planetesimal belt, we are able to place an estimate on the maximum rates that planets could possibly scatter material inwards.

\begin{figure}
\includegraphics[width=0.48\textwidth]{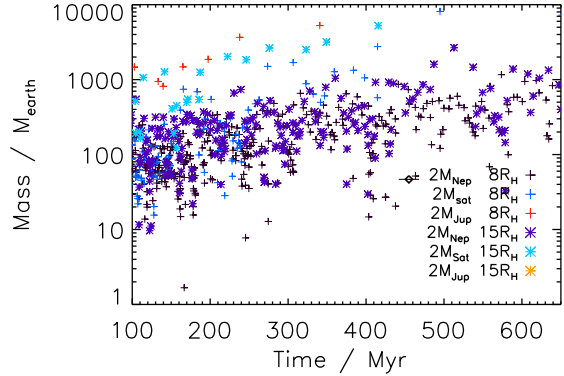}
\caption{The total outer belt mass required if scattering by a chain of equal mass planets, as shown in Fig.~\ref{fig:multi}, or detailed in Table~\ref{tab:multi}, is to replenish an exozodi inside of 1AU around Vega at the required rate of $10^{-9}M_\oplus/$yr. Each cross represents an individual scattering event shown in Fig.~\ref{fig:multi} and scaled by a factor of $\left (\frac{2.157 M_\oplus}{1M_\oplus}\right)^{1/2} $ to account for Vega's stellar mass. Vega is plotted by the diamond at its age of $455\pm13$ Myr \citep{Yoon2010} and with an outer belt mass of $46.7M_\oplus$, as found by the modelling of \citet{muller}. No error bars are quoted on the mass, however, its model dependence should be noted.   }
\label{fig:vega}
\end{figure}

 The position of the outer belt was already chosen to fit with that of the outer belt in the Vega system. Thus, in order to apply the results of these simulations to Vega, it is only necessary to scale the scattering times by a factor of $\frac{1}{\sqrt{2.157}}\sim 0.68 $ in order to account for the increased stellar mass. These results now represent systems with chains of equal mass planets still separated by $\Delta=8R_H$ or 15$R_H$, but with increased masses of $2M_{Nep}$, $2M_{sat}$ or $2M_J$. This is detailed in Table~\ref{tab:multi}. In Fig.~\ref{fig:vega} we calculate the total mass in the outer belt if these chains of equal mass planets were to scatter material in to the position of the exozodi, which we take to be 1AU, at the required rate of $10^{-9}M_\oplus/$yr. This is the minimum required mass in the outer belt, assuming a constant surface density of $\Sigma(r) dr \propto r^{-1} dr $, between 62 and 120AU. %As was seen in Fig.~\ref{fig:multi} the scattering rates decrease steeply with time. This means that the minimum required mass increases with time, as it becomes increasing difficult to retain high scattering rates at late times. 

Fig.~\ref{fig:vega} clearly shows that by the age of Vega ($455\pm13$ Myr) a very massive outer belt is required in order to sustain the high levels of dust observed. In general the required masses are higher than the observationally derived mass for the belt (shown by the diamond), or even the pre-collisional evolution initial mass of $55.5M_\oplus$ \citep{muller}. Given that the plotted values assume that all material scattered inwards is converted to the observed small dust grains and thus are likely to underestimate the belt mass required, this places doubt on the ability of Vega's outer belt to replenish the exozodi. Such a conclusion, should, however, be viewed in the light of the large uncertainties, in particular on the rate at which Vega's exozodi needs to be replenished. A small decrease from the currently used value of $10^{-9}M_\oplus/$yr, would mean that a chain of low mass planets could readily scatter material at the required rates. Another possibility is that the scattering process in Vega started less than 455Myr ago and therefore that the scattering rates are higher at the current epoch. To summarise, the uncertainties on the values of various parameters are sufficiently large that these simulations by no means rule out steady-state scattering by a chain of planets to replenish Vega's exozodi, however, the evidence presented in favour of such scattering requires the existence of many closely spaced, terrestrial to Saturn mass planets, orbiting between the two belts.

%%%%%%%%%%%%%%%%%%%%%%%%%%%%%%%%%%

%%%%%%%%%%%%%%%%%%%%%%%%%%%%%%%%%%

\subsubsection{ $\eta$ Corvi}
\label{sec:eta}
 $\eta$ Corvi as a significantly older ($1.4\pm0.3$Gyr) system, with an extremely high infrared excess. Observations of $\eta$ Corvi in the mid-infrared and sub-mm suggest that it has both a cold and a warm component  \citep{resolveHD69830, Chen06, wyatt_etacorvi}.  Both \cite{resolveHD69830} and \cite{Lisse12} find that the warm component is inside of 3.5AU and \cite{Lisse12} find that it can be explained by $9 \times 10^{18}$kg $\sim 10^{-6}M_\oplus$ of dust in particles between $0.1$ and $100\mu$m in size, with a lifetime of 3-6 yrs against collisions. This suggests that this dust needs to be replenished at a rate of $\sim 10^{-6}M_\oplus/$yr in order to retain the current observed levels. Sub-mm observations of the outer belt, modelled in \cite{wyatt_etacorvi} find a radius of $150\pm20$AU and a mass of $\sim 20M_\oplus$. 

Again the scattering rates shown in Fig.~\ref{fig:multi} can be applied to $\eta$ Corvi, hypothesising the existence of chains of equal mass planets orbiting interior to the outer belt. Again, the planetary system of $\eta$ Corvi may differ radically from such chains of equal mass planets, nonetheless, such systems provide a useful upper limit on the possible scattering rates. In this case, the scattering rates determined in Fig.~\ref{fig:multi} apply at times greater by a factor $\left(\frac{130 AU}{62 AU}\right)^{3/2}\left(\frac{1M_\odot}{1.4M_\odot}\right)^{1/2}\sim 2.57$, to account for the differences in belt radius and stellar mass. The scattering rates now apply to chains of planets with slightly higher masses, scaled upwards by a factor of 1.4, as detailed in Table~\ref{tab:multi}. The scattering rates now apply to a system with an exozodi at 2AU, slightly inside the observed position of less than $3.5$AU, however, this could only potentially cause a small change in the calculated scattering rates.

\begin{figure}
\includegraphics[width=0.48\textwidth]{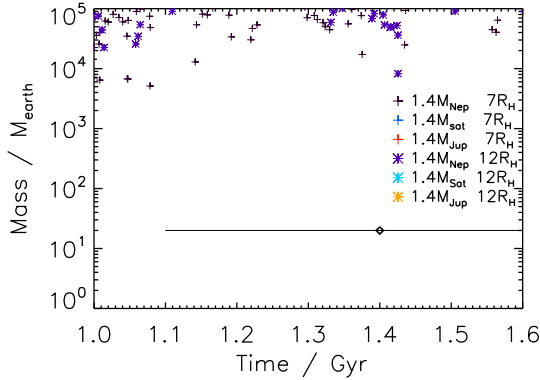}
\caption{The same as Fig.~\ref{fig:vega}, but for $\eta$ Corvi, with its required scattering rate of $\sim 10^{-6}M_\oplus/$yr. The scaling factor this time is $(\frac{130 AU}{62 AU})^{3/2}(\frac{1M_\odot}{1.4M_\odot})^{1/2}$, to account for the difference in stellar mass and radius of the outer belt. $\eta$ Corvi is plotted by the cross, at its age of $1.4\pm0.3$Gyr and with an outer belt mass of $\sim20M_\oplus$ \citep{wyatt_etacorvi}. Again, no error bars are quoted on the mass of the outer belt, but its model dependence should be noted.   }
\label{fig:eta}
\end{figure}

The level of excess observed in this system, and therefore the replenishment rate required, is significantly higher than for Vega's exozodi. This means that the minimum required outer belt mass to replenish the exozodi at $\sim 10^{-6}M_\oplus/$yr, shown in Fig.~\ref{fig:eta}, are ridiculously high. This is clear evidence that the warm dust observed in $\eta$ Corvi is at levels far too high for it to be resupplied by the scattering of material inwards from the observed outer planetesimal belt by a chain of stable planets. Another explanation is required for this system. %These simulations show that if the hot dust could still be resupplied by the outer belt if we observing the system within around a million years of an event that moved a planet into the outer belt, either by scattering or migration. This is a significantly more plausible explanation for this system. 

%%%%%%%%%%%%%%%%%%%%%%%%%%%%%%%%%%

%%%%%%%%%%%%%%%%%%%%%%%%%%%%%%%%%%

\section{Discussion}
\label{sec:discussion}

In this work we have considered some example planetary systems in order to investigate the hypothesis that high levels of exozodiacal dust observed in some planetary systems could be retained by material scattered inwards from an outer planetesimal belt. We used N-body simulations to determine the scattering rates from chains of equal mass planets, on circular, co-planar orbits, interior to an outer planetesimal belt. These were used to assess the validity of this hypothesis by comparison with observations. We used our simulations to outline the properties (mass and radius) of debris discs that are capable of scattering material inwards at sufficiently high rates, which were then compared with our ability to detect such discs and whether the masses are consistent with collisional evolution. In general, this is highly dependent on the properties of individual systems and thus, the most definitive conclusions can only be made by comparison with observed systems, as was done here for Vega and $\eta$ Corvi. We now discuss some of the uncertainties that should be taken into account when considering the results presented in this work.

\subsection{Uncertainties}
The biggest uncertainty in these comparisons is the determination of the rate at which the exozodi must be supplied with material in order to retain the observed levels of exozodiacal dust. This is calculated from a fairly uncertain mass of dust in the exozodi, combined with an even more uncertain timescale on which the dust is removed. The mass calculated for the dust is very model dependent, with a particular dependence on the size distribution and size range of grains considered. For Vega's exozodi, \cite{Defrere_Vega} create a grid of model discs, varying the composition, optical properties, radial and size distribution of the dust, and then use Bayesian statistics to assess which model provides the best fit to the observations. The largest grain size they consider is 1mm, but this upper limit is poorly constrained. The disc will be removed either by radiative forces, on dynamical timescales  of $\sim$yr, or on collisional timescales, also $\sim$yr. Both these timescales depend on the exact properties of the disc and are fairly uncertain. The radial location of the dust is also not precisely constrained, however, this does not significantly effect the scattering rates. 

Similar uncertainties, although less critical in terms of the current hypothesis, also apply to our knowledge of the properties of the outer disc. For the systems where the outer belt is resolved, the radial location of the dust belt is fairly well constrained. The total mass of the disc, however, depends on the extrapolation of the observed population of small dust particles to larger sizes. Although model dependant, detailed numerical modelling  \citep[e.g.][]{Thebault03, Thebault07, Krivov06, lohne} finds that the ratio of the total mass of the disc to the observed dust mass calculated using realistic size distributions does not differ significantly from simplistic estimates made using the size distribution from a steady state collisional cascade ($n(D) dD \propto D^{7/2}dD$ \citep{Dohnanyi} with a power index of $\frac{7}{2}$ \citep{Tanaka96}). Thus, the calculated values for the mass should not vary too dramatically. The biggest uncertainty is in the size of the largest body in the collisional cascade, which can alter the calculated mass significantly. It is also possible that larger bodies exist that are not yet in collisional equilibrium. These could contribute to the scattering process, but only in a stochastic manner. Given our total lack of information regarding such bodies, it seems unreasonable to invoke their contribution to the scattering rates, although, evidence exists in some systems that such stochastic collisions may produce the observed exozodi \citep{Lisse2009, Lisse12}.

The second biggest uncertainty in our calculations regards the conversion of the bodies scattered inwards from the outer belt, with a size distribution reflecting that of the outer belt, to the observed small dust. In order to obtain maximum possible scattering rates, we considered that this process is 100\% efficient. There are three main processes by which larger bodies are converted to small dust, collisions, evaporation or spontaneous disruption. Marboeuf et al, in prep investigated in detail the evaporation of comets as a potential mechanism to produce exozodiacal dust. They find that it is reasonably efficient in high luminosity systems, with a maximum possible efficiency that reflects the dust:ice ratio in the comets (generally about 1, so 50\% efficient). Dust may also be produced as larger bodies are ground down by collisions, or cometary bodies may spontaneously disrupt, as suggested as an explanation for the Solar System's zodiacal cloud \citep{Nesvorny10}. None of these processes will be 100\% efficient at converting large bodies to the observed small grains. In this work, we merely note that where outer belt masses have been calculated, these are very much minimum values, rather than attempting to make any conclusions regarding the conversion of the scattered bodies to the observed small dust. 

%Collisions may occur between several bodies on eccentric orbits, or between bodies scattered into an already formed dust disc. In both cases it is unlikely that all of the mass is converted to small dust in the exozodical dust disc, if merely due to the conservation of angular momentum. \cite{Nesvorny10} suggests that the dust in the Solar System's zodiacal cloud was supplied by the spontaneous disruption of Jupiter Family comets, a phenomenum that has been observed for comets in our Solar System \citep{Weissman1980} It was not the intention of this work to investigate any of these processes in detail. We leave them as the subject of further work, but note that the calculated total mass required in the outer belt is a minimum. 

There are then several uncertainties based on the initial conditions used in our simulations, which were discussed in detail in Sec.~\ref{sec:initialconditions}. It is, however, worth reiterating a dependence of our results, on the fact that planetesimals remain on orbits such that they can be scattered by the planet, at the end of planet formation. Should this prove not to be the case, whether due to planetary migration, or an inaccurate estimation of the parameter that we used to characterise this clearing, $a_{min}$, our scattering rates will only decrease. Again, reiterating the fact that the outer planetesimal belt masses determined by our simulations are very much minimum values.  We also assumed that all planets were fully formed at $t=0$, which of course is not generally true, particularly for terrestrial mass planets. During terrestrial planet formation, the scattering processes are likely to be similar to those seen in our simulations, albeit controlled by the somewhat smaller planetary core. Thus, we do not anticipate that our results would be significantly altered if planet formation was only completed some time after $t=0$.

Of potential concern is how representative our example systems are of the full range of real planetary systems. Such well ordered planetary systems are unlikely to be found (often) in nature. In particular, it is unlikely that planet formation would have been equally efficient throughout any planetary system. They should, however, 
 be representative of the range of scattering rates possible from the full range of real planetary systems, as a function of time. This hypothesis is supported by test simulations of two further planetary systems, with random architectures, which were found to scatter material inwards at rates that lie neatly within the range of scattering rates found for our example systems.  As discussed, for example in Sec.~\ref{sec:scaling} or Sec.~\ref{sec:multi}, it is easy to design planetary systems with `blockages' where the scattering rates tend to zero. However, given our understanding of the dependence of the scattering rates on the planet masses and separations (see discussion in Sec.~\ref{sec:scaling}), we anticipate that very few systems can scatter material inwards after millions of years of evolution, at rates higher than those outlined here for a chain of tightly packed Neptune mass planets (given constraints on the outer belt mass, stellar mass, age {\it etc}). It is, however, still possible that a few very specific architectures exist where the scattering rates are enhanced beyond those seen in our simulations, for example due to resonant interactions. Alternatively systems with eccentric planets or planetesimals may prove to scatter material inwards at higher rates, although test simulations suggest that they can only do so on short timescales.

There are two other potential contributors to the scattering rates, that we have ignored so far. Both contribute at such low levels that we feel justified in leaving them out of our calculated rates. The first is particles that started on orbits between the planets. If the planets are sufficiently widely separated, it may be that some material was left orbiting between the planets, rather than being accreted onto the planets. In general such planetesimals are scattered on short timescales, however, if the separation of the planets is larger than their dynamical influence, \ie $\Delta R_H >  \delta a_{chaos}$, planetesimals may remain on stable orbits between the planets, even on long timescales. The scattering process for such planetesimals is very similar to that for the outer planet in our simulations, except on shorter timescales, given the shorter orbital periods. Thus, at late times the contribution of such particles is likely to be small. The only question therefore is whether such belts of planetesimals, or indeed planetesimals in the process of being scattered inwards, may be detectable.

 The second potential contribution is from particles that started far outside of the outer planet's chaotic zone and are scattered on long timescales. In a multi-planet system, the dynamics are more complicated and there are a range of secular and multi-body resonances that can influence particles in the belt. These have been shown to provide a small flux of material into Neptune's chaotic zone in our Solar System \citep[e.g.][]{duncan95,Holman93, LevisonDuncan97, Emelyanenko04, Morbi97}. This contribution cannot be significant as in our simulations to test this effect with a particle density of 50 particles per AU, not a single particle, that started outside of $a_{max}= a_{pl} + 2\delta a_{chaos}$ and inside of 100AU, was scattered inwards, although there was evidence of the effect of the inner planets on the outer regions of the belt.  %In this work... ?? SIMULATIONS ARE RUNNING!  

\subsection{Comparison with real planetary systems}

All of the potential uncertainties discussed in the preceding section need to be taken into account when comparing the results of our simulations with observed exozodi. In Fig.~\ref{fig:mass_detlim} we determined the mass and radius required for the outer belt, such that it can scatter material to the location of the exozodi at a sufficiently high rate. These requirements could change significantly, for example, if our estimate on the rate at which material needs to be scattered inwards was wrong, or the conversion of the scattered large bodies to small dust is inefficient. Changes of several orders of magnitude are already seen between the example systems considered here. Given a diversity of planetary systems, this would naturally result in a diverse population of exozodi, that could help to explain the range of exozodi detections and non-detections found so far. These uncertainties render insignificant the possible differences in the scattering rates caused by the unrealistic assumption that the radial location of the exozodi scales with that of the outer belt. 

 A general conclusion from Fig.~\ref{fig:mass_detlim} is that the longer lifetimes and scattering timescales of large radii outer belts are required to sustain high scattering rates on long timescales. Unfortunately the population of large radii debris discs is the least well constrained, although sub-mm instruments such as ALMA and SCUBA2 may shed light on this in the near future. For a more detailed comparison, we considered two example systems, Vega (Sec.~\ref{sec:vega}) and $\eta$ Corvi (Sec.~\ref{sec:eta}). For $\eta$ Corvi, the mass of observed outer belt is orders of magnitude too low to retain the exozodi at its currently observed levels and it therefore seems more likely that there is another explanation for this system. For Vega, the replenishment of the exozodi by a chain of planets seems unlikely, but given the uncertainties discussed above, it is not ruled out, although very specific constraints on the system would be required.

Our N-body simulations necessarily only consider a very limited range of all possible planetary system architectures. We consider these to be a fair representation of the range of scattering rates possible in any arbitrary planetary system, and discussed their application to any planetary system in detail in Sec.~\ref{sec:scaling}. However, it still remains possible that we have missed some very specific architecture involving unequally spaced planets of different masses, potentially on eccentric or inclined orbits, that is significantly more efficient at scattering material, for example where secular or mean motion resonances enhance the scattering rates. However, within the limitations of reasonable computing timescales, a full investigation was not possible in this work.

%We feel that the range of scattering rates calculated in our simulations are a fair representation of the range of scattering rates possible in any arbitrary planetary system, where the planets orbit on circular orbits. The scattering rates for eccentric or inclined planets may well be significantly higher, but it is beyond the scope of this work to characterise them. We have only considered chains of equal mass planets. These are unlikely to ever form since planet formation is unlikely to ever be equally efficient throughout the system. However, the scattering rates in any system will be dominated by the outer planet and then either hindered or enhanced by the interior planets. Thus, in general, the highest possible scattering rates at late times for a chain of planets, are those for a chain of Neptune mass planets. To calculate many of the properties of potential outer belts, we scaled the scattering rates with the radius of the outer belt. This scaling is fairly robust, although it assumes equivalent changes in the position of the exozodi. Thus, the scattering rates are potentially slightly over estimated for larger radii debris discs and underestimated for smaller radii discs, for example the scattering rate of a belt at 240AU only considers particles scattered inside of 3AU. 

\subsection{Application to unstable systems}
The results presented in this work find high scattering rates for the first thousands or millions of years of a system's evolution, but much lower rates in stable planetary systems after hundreds of millions of years. However, there is a potential way to get high, albeit temporary, scattering rates in such older mature systems. This could happen if the empty chaotic region surrounding a planet is refilled with fresh material. A way to achieve this is if the planet is displaced from its current orbit by a dynamical instability and scattered into a planetesimal belt, or if it migrates into such a planetesimal belt at late times. Such processes could restart the scattering processes and significantly increase the scattering rates. From our simulations, we can estimate that the scattering rates could remain high, even for thousands to millions of years after such an event. However, this is only an estimate because we did not consider any test particles closer to the planet than $a_{min}=a_{pl} + \frac{\delta a_{chaos}}{2}$ and if the chaotic zone were refilled with planetesimals after an instability, it is likely that the planetesimal belt will be cold. Further detailed work is required to investigate this hypothesis.

\section{Conclusions}
\label{sec:conc}

In this work we have considered the conundrum that high levels of exozodiacal dust are observed in some systems, despite it being clear that such small dust has very short lifetimes against collisions or radiative forces. We considered the hypothesis that the exozodiacal dust could be replenished by material from an outer belt, scattered inwards by planets. We presented N-body simulations that calculate the efficiency of scattering by a chain of equal mass planets, in a way that can be applied to a wide range of different planetary systems, irrespective of whether or not exozodiacal dust has been detected. The outer planet scatters planetesimals as it sculpts the inner edge of the planetesimal belt. Particles are cleared from the chaotic zone on relatively short timescales, but these particles may continue to be scattered multiple times by the planet, evolving at constant pericentre or apocentre, until they are scattered sufficiently far inwards that they are passed onto an interior planet or ejected. In this manner particles can be passed along a chain of planets and may only reach the location of the exozodi, some after hundreds of millions of years. 

Our simulations show that it is possible for this scattering to be at a sufficiently high rate that it could retain the currently observed levels of exozodiacal dust. This, however, places very stringent constraints on the architecture of the planetary system, namely that it contain a chain of closely spaced, low-mass planets orbiting interior to a massive, large radii outer planetesimal belt. Without a complete census of the architecture of planetary systems and/or detailed statistics about the exozodi population, it is difficult to make definitive conclusions about the proportion of systems in which this scenario could apply. We considered in detail two systems with observed outer belts and exozodi, and we hypothesise that there may be as yet undetected planets orbiting between the observed exozodi and outer belt. We find that for $\eta$ Corvi, an old, sun-like star with very high levels of infrared excess, that the exozodi cannot be retained at its current levels, on Gyr timescales, with this process, whereas for Vega a very contrived architecture would be required for the planetary system to retain the levels of exozodiacal dust observed around Vega. 

In general in this work we show that material is scattered into the inner systems of many planetary systems, at rates that depend on the architecture of the planetary system, but are in general low, particularly after hundreds of millions of years. We put this forward as an explanation for the observations of high levels of exozodiacal dust, but find that sufficient material is only scattered inwards in systems with very contrived architectures. An alternative solution for, particularly the older planetary systems with exozodiacal dust, could be that we are observing these systems in the aftermath, albeit the long aftermath, of a dynamical instability that led to a planet being scattered into the outer belt.

\section{Acknowledgements}
We thank Dimitri Veras, Herv\'{e} Beust, Virginie Faramaz, J\'{e}remey L\'{e}breton, Julien Vanderportal, Quentin Kral and Olivier Absil for discussions that contributed to this work. We also thank the anonymous referee for their comments.  
We acknowledge the support of the ANR-2010 BLAN-0505-01 (EXOZODI).
Computations presented in this paper were performed at the Service Commun de Calcul Intensif de l'Observatoire de Grenoble (SCCI) on the super-computer funded by Agence Nationale pour la Recherche under contracts ANR-07-BLAN-0221, ANR-2010-JCJC-0504-01 and ANR-2010-JCJC-0501-01. 

\bibliographystyle{aa} % style aa.bst

\bibliography{ref}

\end{document}